\documentclass[lettersize,journal]{IEEEtran}
\pdfoutput=1 
\usepackage{amsmath,amsfonts}
\usepackage{algorithmic}
\usepackage{algorithm}
\usepackage{array}
\usepackage[caption=false,font=normalsize,labelfont=sf,textfont=sf]{subfig}
\usepackage{textcomp}
\usepackage{stfloats}
\usepackage{url}
\usepackage{verbatim}
\usepackage{graphicx}
\usepackage{cite}
\usepackage{bm}

\graphicspath{{figures/}}
\begin{document}
\title{Sparse Bayesian Learning-Based 3D Spectrum Environment Map Construction —— Sampling Optimization, Scenario-Dependent Dictionary Construction and Sparse Recovery}

\author{Jie Wang, Qiuming Zhu,~\IEEEmembership{Member,~IEEE,} Zhipeng Lin,~\IEEEmembership{Member,~IEEE,} Qihui Wu,~\IEEEmembership{Senior Member,~IEEE,} Yang Huang,~\IEEEmembership{Member,~IEEE,} Xuezhao Cai, Weizhi Zhong,~\IEEEmembership{Member,~IEEE,} Yi Zhao
\thanks{This work was supported in part by the National Key Scientific Instrument and Equipment Development Project (No. 61827801), in part by the National Natural Science Foundation of China (No. 62271250), in part by Natural Science Foundation of Jiangsu Province (No. BK20211182), in part by the open research fund of National Mobile Communications Research Laboratory, Southeast University, No. 2022D04.}
\thanks{J. Wang, Q. Zhu, Z. Lin, Y. Huang, Y. Zhao, Q. Wu and X. Cai are with The Key Laboratory of Dynamic Cognitive System of Electromagnetic Spectrum Space, College of Electronic and Information Engineering, Nanjing University of Aeronautics and Astronautics, Nanjing 211106, China  (e-mail: {bx2104906wangjie; zhuqiuming; linlzp; yang.huang.ceie; zhaoyi}@nuaa.edu.cn, wuqihui2014@sina.com, caixz3312@outlook.com).}
\thanks{W. Zhong is with The Key Laboratory of Dynamic Cognitive System of Electromagnetic Spectrum Space, College of Astronautics, Nanjing University of Aeronautics and Astronautics, Nanjing 211106, China (e-mail: zhongwz@nuaa.edu.cn).}
}

\markboth{}%
{Shell \MakeLowercase{\textit{et al.}}: A Sample Article Using IEEEtran.cls for IEEE Journals}

\IEEEpubid{0000--0000/00\$00.00~\copyright~2021 IEEE}

\maketitle

\begin{abstract}
The spectrum environment map (SEM), which can visualize the information of invisible electromagnetic spectrum, is vital for monitoring, management, and security of spectrum resources in cognitive radio (CR) networks. In view of a limited number of spectrum sensors and constrained sampling time, this paper presents a new three-dimensional (3D) SEM construction scheme based on sparse Bayesian learning (SBL). Firstly, we construct a scenario-dependent channel dictionary matrix by considering the propagation characteristic of the interested scenario. To improve sampling efficiency, a maximum mutual information (MMI)-based optimization algorithm is developed for the layout of sampling sensors. Then, a maximum and minimum distance (MMD) clustering-based SBL algorithm is proposed to recover the spectrum data at the unsampled positions and construct the whole 3D SEM. We finally use the simulation data of  the campus scenario to construct the 3D SEMs and compare the proposed method with the state-of-the-art. The recovery performance and the impact of different sparsity on the constructed SEMs are also analyzed. Numerical results show that the proposed scheme can reduce the required spectrum sensor number and has higher accuracy under the low sampling rate.

\end{abstract}


\section{Introduction}
\IEEEPARstart{Wi}{th} the increase of various electronic devices, i.e., radio, radar, navigation and so on, the electromagnetic environment becomes significantly complex \cite{Ding18CM, Ahmad15CST, Wang11JSTSP}. The Spectrum Environment Map (SEM) visualizes the spectrum related information, including the time, frequency and the received signal strength (RSS) of signals, and the locations of the sensor devices, on a geographical map \cite{Yilmaz13CM, 16CST}. It  is useful for the abnormal spectral activity detection, radiation source localization, radio frequency (RF) resource management, and so on. However, constructing an accurate SEM for the scenario with lots of buildings is still difficult. This is because the sensor device number and sampling time are limited in practice. Besides, with the development of space-air-ground integrated communication networks, electronic devices are distributed in the three-dimensional (3D) space. \IEEEpubidadjcol

Many SEM reconstruction methods have been proposed recently \cite{Fujii17TCCN, HU20TN, DALLANESE11TVT, TANG16Access, Shrestha22TSP, HAN20Sensors, YILMAZ15WCMC, PESKO15EURASIP, SHEN19TVT, Shen22TWC,DEMO22}. They can be divided into two major categories, i.e., the direct construction methods driven by massive measured data and the indirect construction methods driven by channel propagation characteristics. 

The direct construction methods typically employ interpolation algorithms to recovery the missing data by mining the correlation between the given measured data. In \cite{Sato21TVT}, the authors utilized the Kriging interpolation to construct the two-dimensional (2D) SEM at different frequencies. A tensor completion method was used in \cite{TANG16Access} to recovery the missing data in both the spatial domain and temporal domain. Machine learning techniques have also been adopted for the data-driven REM construction. For example, from the perspective of image processing, the authors in \cite{HAN20Sensors} developed a generative adversarial network (GAN) to obtain the SEM based on the sampling data. Deep neural networks were used in \cite{Shrestha22TSP} to "learn" the intricate underlying structure from the given data and constructed the SEM. Nevertheless, the aforementioned methods mainly focus on the 2D SEM re-construction and can only achieve satisfactory performance by using a large amount of sampling data. 

The indirect construction methods can greatly reduce the number of sampling data by using the rule of wireless signal propagation \cite{maokai22}. In \cite{YILMAZ15WCMC,  PESKO15EURASIP}, the limited sampling data was used to estimate the transmitters’ information, and then the missing data was recovered by using the ideal propagation model. Considering the realistic propagation model, the authors in \cite{Lee17TWC} modeled the spectrum data as the tomographic accumulation of spatial loss field. 

However, in many practical cases, the number of sampling sensors is very limited and the spatial sampling rate may be much lower than the Nyquist rate. To tackle this issue, the compressed sensing (CS) technique is applied, which decomposes the measurement into the linear superposition of the sensing matrix and the sparse signal to realize data recovery under sparse sampling.  For example, the authors in \cite{Bazerque10TSP} constructed the SEM based on least absolute shrinkage and selection operator (LASSO) with random sparse sensing data. In \cite{Shen22TWC}, the authors proposed an improved orthogonal matching pursuit algorithm (OMP) to build compressed SEM with right-triangular (QR) pivoting based sampling locations optimization. However, the traditional CS approaches, such as LASSO \cite{Tibshirani96JRS}, OMP \cite{Tropp07TIT} and linear programming \cite{Bertsimas97ILO}, are point estimation for sparse signals. Besides, the SEM sensing matrix usually has high spatial correlation, which would greatly deteriorate the recovery performance from the noisy measurements. The sparse Bayesian learning (SBL) \cite{Tipping01JMLR, Tipping00UAI, Kiaee16TNNLS} can recover the exact sparse signal under the high correlated sensing matrix. A SBL-based SEM construction algorithm was proposed in \cite{Huang15TVT}.  The authors used a Laplacian function to describe the signal propagation model for sparse dictionary construction, which does not consider the realistic propagation scenario, especially the impact of buildings. Furthermore, the authors arranged the sampling sensors randomly without considering the sampling location (SL) optimization for different scenarios.    

To fill these gaps, we propose a novel 3D SBL-based SEM construction scheme with optimized sampling positions of sensors and  a scenario-dependent dictionary with full consideration of scenario characteristics. The main novelties and contributions of this paper are summarized as follows:
\begin{itemize}	
	\item  A channel propagation dictionary design method for SEM construction under sparse sampling is proposed. Combined with the ray tracing (RT) simulation technology and the spatial interpolation algorithm, the scenario-dependent SEM construction dictionary is obtained.
	
	\item  A maximum mutual information (MMI)-based measurement matrix optimization architecture is proposed. By modeling the construction problem as a communication channel model, the objective function is derived from the SBL framework and solved by the greedy algorithm, which  improving the efficiency of spectrum data acquisition.
	
	\item  An improved SBL algorithm based on cluster analysis is developed for 3D compressed spectrum recovery. The maximum and minimum distance (MMD) clustering and dynamic threshold pruning are combined with the SBL, which can recover sparse signals and achieve accurate SEM construction.	
\end{itemize}

The rest of this paper is organized as follows. Section II gives the 3D SEM construction model and the sparse sampling model. In Section III, the details of proposed 3D SEM construction scheme is given and demonstrated. Then, Section IV presents the simulation and comparison results and Section V gives some conclusions.

\begin{figure}[!t]
	\centering
	\includegraphics[width=3in]{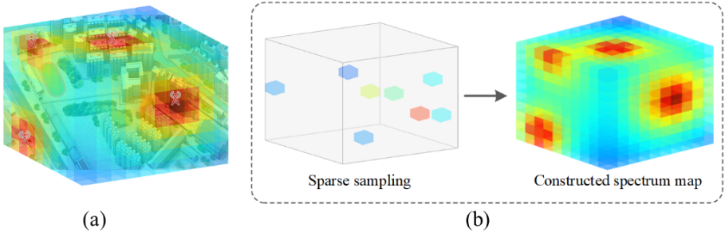}
	\caption{(a): Realistic continuous 3D spectrum map under an urban scenario; (b): Constructed discrete 3D spectrum map.}
	\label{fig1}
\end{figure}

\begin{figure}[!t]
	\centering
	\includegraphics[width=2.5in]{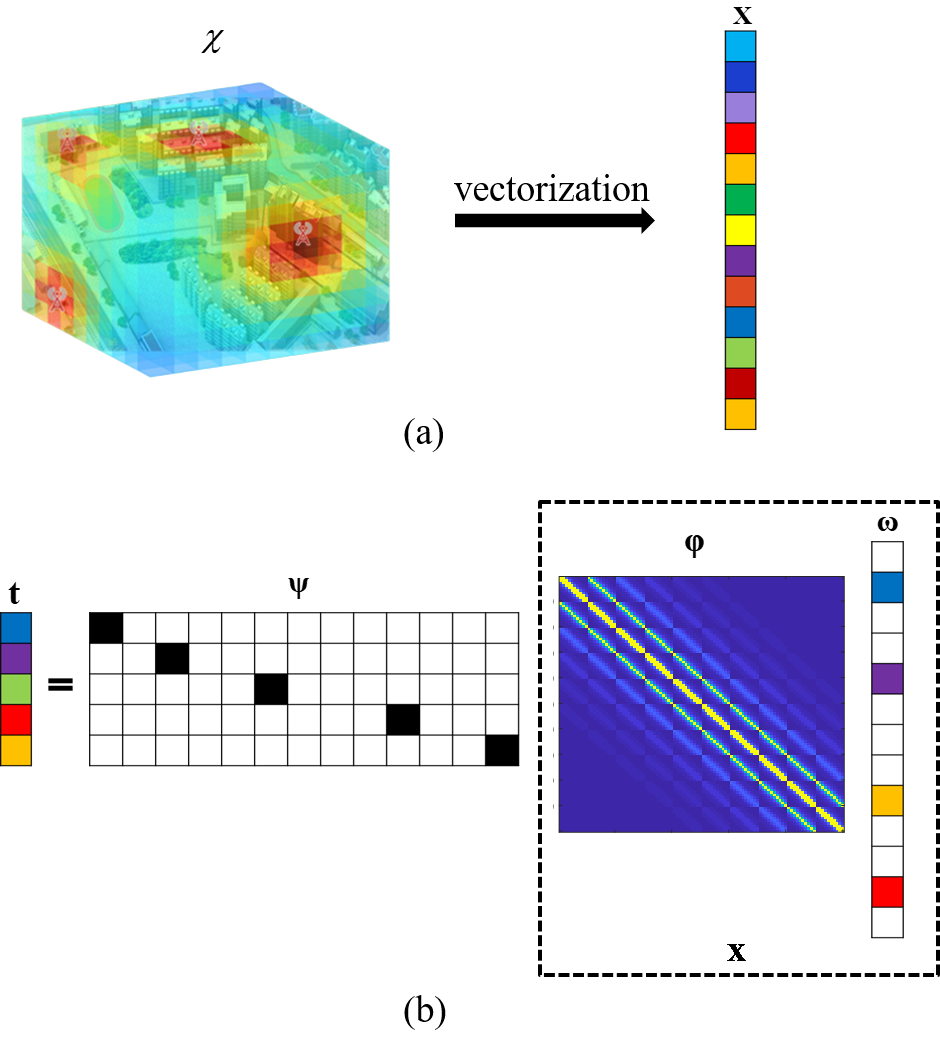}
	\caption{Schematic illustration of 3D SEM construction process based on sparse sampling.}
	\label{fig2}
\end{figure}

\section{System Model}
\label{SYSTEM}
\subsection{3D SEM Model}
As shown in Fig.\ref{fig1}, the region of interest (ROI) is discretized into several small cubes. Each cube is colored according to its RSS, where red cubes represent high RSS values and blue cubes represent low RSS values. The ROI constitutes a spectrum tensor $\bm{\chi}  \in {\Re ^{{N_x} \times {N_y} \times {N_z}}}$ in the 3D space, where ${N_x}$, ${N_y}$, ${N_z}$ indicate the grid number along $x$, $y$, $z$ dimensions, respectively. Technically, 3D SEM construction in this paper aims to recovery RSS values of all $N = {N_x} \times {N_y} \times {N_z}$ cubes based on the known RSS values of sampling cubes.

A sparse signal vector $\bm{\omega}  = {\left[ {{\omega _1},{\omega _2}, \ldots ,{\omega _n}, \ldots ,{\omega _N}} \right]^{\text{T}}} \in {\mathbb{R}^{N \times 1}}$ can be defined as
\begin{equation}
{\omega _n} = \begin{cases}
P_n^t,&{\text{if there is a RF transmitter in the }}n{\text{th cube,}} \\ 
{0,}&{\text{otherwise.}} 
\end{cases}
\label{eq1}
\end{equation}
where $P_n^t$ is the transmitting power of the $n$th RF transmitter. \bm{$\omega$} is a $K$-sparse signal vector with $\left\| \bm{\omega}  \right\|_0^{} = K$. That is if there are $K$ stationary RF transmitters denoted as $\left\{ {{{\mathbf{T}}_k}} \right\}_{k = 1}^K$, where ${{\mathbf{T}}_k} = \left( {x_k^t,y_k^t,z_k^t} \right)$ is the location of $k$th RF transmitter, we have $K\left( {K \ll N} \right)$ nonzero elements in \bm{$\omega$}.

Suppose that we select $M$ SLs from all $N$ cubes denoted by $\left\{ {{{\mathbf{S}}_m}} \right\}_{m = 1}^M$ and ${{\mathbf{S}}_m} = \left( {x_m^s,y_m^s,z_m^s} \right)$. The sampling rate is $r = M/N$. The Euclidean distance from the $k$th transmitter to the $m$th SL can be written as
\begin{equation}
{d_{m,k}} = {\left\| {{{\mathbf{T}}_k} - {{\mathbf{S}}_m}} \right\|_2}.
\label{eq2}
\end{equation}

Under the line-of-sight (LOS) condition, the RSS from the $k$th transmitter to the $m$th SL can be approximated by using the free space propagation model as
\begin{equation}
P_{m,k}^r = \frac{{{G_t}{G_r}{\lambda ^2}P_k^t}}{{{{\left( {4\pi } \right)}^2}d_{m,k}^\eta }},
\label{eq3}
\end{equation}
where ${G_t}$ and ${G_r}$ are the antenna gain of transceivers, $P_k^t$ denotes the transmitting power, $\lambda $ is the wavelength of carrier, and $\eta $ is the path loss exponent. Since the RSS sensed at each SL may include the receiving power of several RF transmitters \cite{Shen22TWC}, the total RSS can be approximately expressed as  
\begin{equation}
{t_m} = \sum\limits_{k = 1}^K {P_{m,k}^r} .
\label{eq4}
\end{equation}

It should be mentioned that the realistic propagation model is very complex due to reflection, diffraction and other propagation phenomena. The above data recovery method for the unsampled cubes is only suitable for the LOS propagation scenarios. 

\subsection{Sparse Sampling Model}
\label{SSM}
As shown in Fig.\ref{fig2} (a), the spectrum tensor $\bm{\chi}$ is firstly vectorized into ${\mathbf{X}} \in {\mathbb{R}^{N \times 1}}$. Compared with the total number of discretized cubes in the ROI, the number of stationary RF transmitters is much small ($K \ll N$). Since the spectrum vector ${\mathbf{X}}$ has high spatial correlation, it can be represented by the product of a sparse dictionary $\bm{\varphi}  \in {\mathbb{R}^{N \times N}}$ and the sparse signal \bm{$\omega$} as
\begin{equation}
{\mathbf{X}} = \bm{\varphi} \bm{\omega} ,
\label{eq5}
\end{equation}
where the element ${\varphi _{i,j}}$ of sparse dictionary is defined as the channel gain or propagation path loss between the $i$th and the $j$th cubes. Let us set the RSS vector of SLs as $\bm{t} \in {\mathbb{R}^{M \times 1}}$. The measurement matrix $\bm{\psi}  \in {\mathbb{R}^{M \times N}}$ can be defined as
\begin{equation}
{\psi _{i,j}} = \begin{cases}
1,&{\text{if the }}i{\text{th SL is at the }}j{\text{th cube,}} \\ 
{0,}&{\text{otherwise.}} 
\end{cases}
\label{eq6}
\end{equation}
where each row of $\bm{\psi}$ has an element of 1 denoting the SL’s position in the ROI. Then, we can have
\begin{equation}
\bm{t} = \bm{\psi} {\mathbf{X}} + \bm{\varepsilon}  = \bm{\psi} \bm{\varphi} {\bm{\omega }} + \bm{\varepsilon}  = {\mathbf{\Phi }}\bm{\omega }  + \bm{\varepsilon} ,
\label{eq7}
\end{equation}
where $\bm{\varepsilon}  \in {\mathbb{R}^{M \times 1}}$ is the measurement noise that obeys the zero-mean Gaussian distribution with variance ${\sigma ^2}$, and ${\mathbf{\Phi}}$ is a sensing matrix. 

This paper recovers the spectrum data by two steps. Firstly, the sparse signal \bm{$\omega$} is recovered by the noisy measurement vector $\bm{t}$ and the sensing matrix ${\mathbf{\Phi}}$. Then, the spectrum vector ${\mathbf{X}}$ is constructed according to (\ref{eq5}). The sparse signal recovery (SSR) is equivalent to solving the following ${l_0}$-minimization problem as
\begin{equation}
\begin{gathered}
 \bm{\hat \omega}  = \arg \min {\left\| \bm{\omega}  \right\|_0}, \hfill \\
\begin{array}{*{20}{c}}
{{\text{s}}{\text{.t}}{\text{.}}}&{\bm{t} = {\mathbf{\Phi }}\bm{\omega }  + \bm{\varepsilon}} 
\end{array}. \hfill \\ 
\end{gathered} 
\label{eq8}
\end{equation}
The ${l_0}$-minimization problem of (\ref{eq8}) is NP-hard, and can be equivalent to a ${l_1}$-minimization problem as \cite{Baraniuk07SP}
\begin{equation}
\begin{gathered}
\bm{\hat \omega}  = \arg \min {\left\| \bm{\omega}  \right\|_1}, \hfill \\
\begin{array}{*{20}{c}}
{{\text{s}}{\text{.t}}{\text{.}}}&{\bm{t} = {\mathbf{\Phi }}\bm{\omega }  + \bm{\varepsilon}} 
\end{array}. \hfill \\ 
\end{gathered} 
\label{eq9}
\end{equation}
The authors in \cite{Herman10JSTSP} have proved that if ${\mathbf{\Phi}}$ satisfies the condition of restricted isometry property, \bm{$\omega$} can be recovered by applying CS algorithms. In this paper, we focus on using the sparse Bayesian theory to solve the recovery problem of the SEM construction. 

\begin{figure*}[!t]
	\centering
	\includegraphics[width=4.5in]{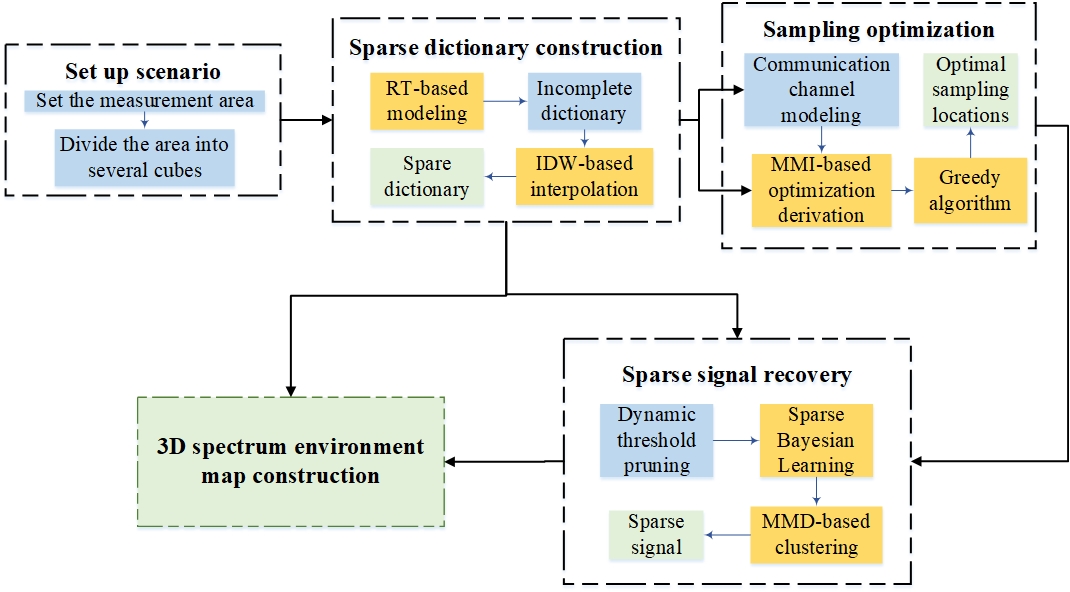}
	\caption{The flowchart of the proposed 3D SEM construction scheme.}
	\label{fig3}
\end{figure*}

\section{3D SEM Construction based on SBL}
\label{SEM}
\subsection{An Overview of the Proposed 3D SEM Construction Scheme}
The proposed 3D SEM construction scheme is illustrated in Fig. \ref{fig3}, which mainly contains the scenario-dependent dictionary construction, the measurement matrix optimization and the 3D SEM construction. Firstly, combined with RT technology and interpolation algorithm, we take the factor of scenario into account and build a scenario-dependent sparse dictionary $\bm{\varphi}$. According to the maximum mutual information criterion, we design the selection scheme of SLs and obtain the measurement matrix $\bm{\psi}$. Then, based on the sparse dictionary and measurement matrix, the sparse signal \bm{$\omega$} can be recovered based on the SBL algorithm. Finally, we utilize the sparse dictionary and sparse signal to construct the full SEM. 

In order to recover the sparse signal \bm{$\omega$} for the SEM construction, we adopt the SBL theory and hierarchical sparse probabilistic model as follows.

1) Noise Model: The sparse regression model (\ref{eq7}) is defined in a zero-mean Gaussian noise $\bm{\varepsilon}$ with unknown variance ${\sigma ^2}$, and then the Gaussian likelihood of $\bm{t}$ can be written as
\begin{equation}
p\left( {\bm{t}|\bm{\omega} ,{\sigma ^2}} \right) = {\left( {2\pi {\sigma ^2}} \right)^{ - M/2}}\exp \left\{ { - \frac{{{{\left\| {\bm{t} - {\mathbf{\Phi }}\bm{\omega} } \right\|}^2}}}{{2{\sigma ^2}}}} \right\}.
\label{eq10}
\end{equation}
A Gamma distribution is then posed on $\beta \left( {\beta  = \left( {\sigma _{}^2} \right)_{}^{ - 1}} \right)$, which is a conjugate prior to the Gaussian distribution and can greatly simplifies the analysis \cite{Babacan10TIP}, as 
\begin{equation}
p\left( {\beta ;c,d} \right) = {\text{Gamma}}\left( {\beta |c,d} \right),
\label{eq11}
\end{equation}
with
\begin{equation}
{\text{Gamma}}\left( {\beta |c,d} \right) = \Gamma {(c)^{ - 1}}{d^c}{\beta ^{c - 1}}{e^{ - d\beta }},
\label{eq12}
\end{equation}
where $c > 0$ is the shape parameter, and $d > 0$ is the scale parameter, $\Gamma (c) = \int_0^\infty  {{t^{c - 1}}{e^{ - t}}dt}$. 

\begin{figure}[!t]
	\centering
	\includegraphics[width=3.5in]{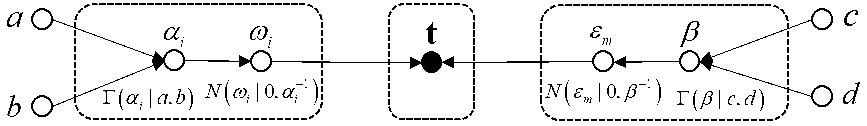}
	\caption{Graphical representation of the SBL model.}
	\label{fig4}
\end{figure}

2) Hierarchical Sparse Prior Model: To induce the sparsity of \bm{$\omega$}, we deploy a sparseness-promoting prior on it. In the Relevance Vector Machine, \bm{$\omega$} is a two-layer hierarchical sparse prior, which shares the same properties with Laplace prior while enabling convenient computation \cite{Tipping01JMLR}. Specifically, each element of \bm{$\omega$} is first posed a zero-mean Gaussian prior
\begin{equation}
p\left( {\bm{\omega} |\bm{\alpha} } \right) = \prod\limits_{i = 1}^N {\mathcal{N}\left( {{\omega _i}|0,\alpha _i^{ - 1}} \right)} ,
\label{eq13}
\end{equation}
where $\bm{\alpha}  = \left[ {{\alpha _1},{\alpha _2}, \ldots ,{\alpha _N}} \right]_{}^{\text{T}}$. Then, to complete the specification of this hierarchical prior, we consider the Gamma hyperpriors over $\bm{\alpha}$ as 
\begin{equation}
p\left( {\bm{\alpha} ;a,b} \right) = \prod\limits_{i = 1}^N {{\text{Gamma}}} \left( {{\alpha _i}|a,b} \right).
\label{eq14}
\end{equation}
The graphical model is shown in Fig.\ref{fig4}. The overall prior $p\left( \bm{\omega}  \right)$ can be obtained by computing the marginal integral of hyper-parameters in $\bm{\alpha}$ as
\begin{equation}
p\left( \bm{\omega}  \right) = \int {p\left( {\bm{\omega} |\bm{\alpha} } \right)p\left( \bm{\alpha}  \right)d} \bm{\alpha} .
\label{eq15}
\end{equation}
Since the integral is computable for Gamma $p\left( \bm{\alpha}  \right)$, the true prior $p\left( \bm{\omega}  \right)$ is a Student-t distribution which can promote sparsity on \bm{$\omega$}\cite{Tipping03ALML}. 

3) Sparse Bayesian Inference: Following the Bayesian inference, the posterior distribution over all unknowns is desired as
\begin{equation}
p\left( {\bm{\omega} ,\bm{\alpha} ,\beta |\bm{t}} \right) \equiv p\left( {\bm{\omega} |\bm{t},\bm{\alpha} ,\beta } \right)p\left( {\bm{\alpha} ,\beta |\bm{t}} \right),
\label{eq16}
\end{equation}
with the decomposition of the ‘weight posterior’ and the ‘hyper-parameter posterior’. It can be inferred that the weight posterior of \bm{$\omega$} is Gaussian
\begin{equation}
\begin{gathered}
p\left( {\bm{\omega} |\bm{t},\bm{\alpha} ,\beta } \right) = \frac{{p\left( {\bm{t}|\bm{\omega} ,\bm{\alpha} ,\beta } \right)p\left( {\bm{\omega} |\bm{\alpha} } \right)}}{{p\left( {\bm{t}|\bm{\alpha} ,\beta } \right)}} \\ 
= \mathcal{N}\left( {\bm{\omega} |\bm{\mu} ,{\mathbf{\Sigma }}} \right), \\ 
\end{gathered} 
\label{eq17}
\end{equation}
with
\begin{equation}
\begin{gathered}
\bm{\mu}  = \beta {\mathbf{\Sigma }}{{\mathbf{\Phi }}^T}{\mathbf{t}}, \hfill \\
{\mathbf{\Sigma }} = {\left( {\beta {{\mathbf{\Phi }}^T}{\mathbf{\Phi }} + {\mathbf{\Lambda }}} \right)^{ - 1}}, \hfill \\ 
\end{gathered} 
\label{eq18}
\end{equation}
where ${\mathbf{\Lambda }} = {\text{diag}}\left[ {{\alpha _1},{\alpha _2}, \ldots ,{\alpha _N}} \right]$. The SBL considers the signal recovery from the perspective of statistics. With sparse prior of \bm{$\omega$} and compressive samples $\bm{t}$, the posteriori probability density function of sparse vector \bm{$\omega$} can be inferred. We further estimate the \bm{$\omega$} with the mean $\bm{\mu}$ and evaluate the accuracy of the recovery by the variance ${\mathbf{\Sigma }}$. To calculate $\bm{\mu}$ and ${\mathbf{\Sigma }}$, we estimate probabilistic model hyperparameters $\bm{\alpha}$ and $\beta $, the details of estimation will be discussed in the Section \ref{SBL}.

\subsection{Scenario-dependent Sparse Dictionary Construction}
The traditional sparse dictionary usually adopts the free-space propagation model without considering the scene information, which is not suitable for urban environments with a lot of buildings. The propagation occurs direct, reflection, and diffraction phenomena in the actual environment. Accordingly, a scenario-dependent dictionary is constructed here by analyzing the characteristics of propagation channel. The RT technique is based on Geometrical Optics and the Uniform Theory of Diffraction. It has been used to predict all the possible propagation path parameters in a given geographic map \cite{Zhu22TVT}. 

Assume that the position of $m$th transmitting cube and $n$th receiving cube are $\bm{m} = [{m_x},{m_y},{m_z}]$ and $\bm{n} = [{n_x},{n_y},{n_z}]$, respectively. The propagation distance of direct path is obtained as 
\begin{equation}
d_{mn}^{} = \sqrt {{{({m_x} - {n_x})}^2} + {{({m_y} - {n_y})}^2} + {{({m_z} - {n_z})}^2}} .
\label{eq19}
\end{equation}
For the indirect path, we define the intersection coordinates of scatterers as $h = ({H_x},{H_y},{H_z})$. Thus, the Euclidean distance of $s$th path ray between $m$ and the scatterer, and the Euclidean distance of $s$th ray between the scatterer and $m$ can be respectively calculated as
\begin{equation}
d_{h,m}^{\text{s}} = \sqrt {{{({H_x} - {m_x})}^2} + {{({H_z} - {m_z})}^2} + {{({H_z} - {m_z})}^2}} ,
\label{eq20}
\end{equation}
\begin{equation}
d_{n,h}^{\text{s}} = \sqrt {{{({H_x} - {n_x})}^2} + {{({H_z} - {n_z})}^2} + {{({H_z} - {n_z})}^2}} .
\label{eq21}
\end{equation}

The proposed RT-based dictionary construction method includes three steps, i.e., decomposition of ray source, tracking rays, and superposition of the filed strength. Firstly, the ray source is decomposed with direct ray, reflection ray and diffraction ray. If the ray arrives at the receiving field position $n$ in LOS propagation, the field intensity of $m$ arriving at $n$ is
\begin{equation}
{{\mathbf{{\mathbf{E}}}}_{{\text{LOS}}}} = E_{1{\text{m}}}^{}\frac{{{e^{ - jld_{mn}^{}}}}}{{d_{mn}^{}}},
\label{eq22}
\end{equation}
where $l$ is the wave number. $E_{1{\text{m}}}^{}$ is the electric field intensity of $1{\text{ m}}$ away from $m$, and $d_{mn}^{}$ is the propagation distance of the direct path in (\ref{eq19}). For the reflected ray, the electric field intensity can be expressed as
\begin{equation}
{\mathbf{E}}_s^{\text{R}} = {{\mathbf{E}}_{{\text{LOS}}}}R\frac{{{{\text{e}}^{ - lj(d_{h,m}^{\text{s}} + d_{n,h}^{\text{s}})}}}}{{d_{h,m}^{\text{s}} + d_{n,h}^{\text{s}}}},
\label{eq23}
\end{equation}
where $R$ is the reflection coefficient. The electric field intensity of the diffraction path can be expressed as
\begin{equation}
{\mathbf{E}}_s^{\text{D}} = \frac{{{{\mathbf{{\mathbf{E}}}}_{{\text{LOS}}}}}}{{d_{n,h}^{\text{s}}}}D\sqrt {\frac{{d_{n,h}^{\text{s}}}}{{d_{h,m}^{\text{s}} \cdot (d_{h,m}^{\text{s}} + d_{n,h}^{\text{s}})}}}  \cdot {{\text{e}}^{ - jl(d_{h,m}^{\text{s}} + d_{n,h}^{\text{s}})}},
\label{eq24}
\end{equation}
where $D$ is the diffraction coefficient. 

Then, according to (\ref{eq22})-(\ref{eq24}), the final received field strength of $n$ can be obtained by vector superposition of all the field strengths
\begin{equation}
\begin{array}{*{20}{c}}
{{{\mathbf{E}}_n} = \sum\limits_{s = 1}^{{N_P}} {{{\mathbf{E}}_s}} ,}&{{{\mathbf{E}}_s} \in \left\{ {{{\mathbf{{\mathbf{E}}}}_{{\text{LOS}}}},{\mathbf{E}}_s^{\text{R}},{\mathbf{E}}_s^{\text{D}}} \right\},} 
\end{array}
\label{eq25}
\end{equation}
where ${N_p}$ is the total number of effective rays. ${{\mathbf{E}}_s}$ is the electric field intensity of direct path, reflection path or diffraction path. It should be mentioned that the direct ray disappears when the link between $m$ and $n$ is blocked. Then the parameter calculation of direct ray can be ignored \cite{Zhu22TVT}. Therefore, the total average RSS of $n$ is
\begin{equation}
p_{}^n = {G^n}{G^m}{\left( {\frac{\lambda }{{4\pi }}} \right)^2}\left| {\frac{{{{\mathbf{E}}_n}}}{{{{\mathbf{E}}_{1{\text{m}}}}}}} \right|{\text{,}}
\label{eq26}
\end{equation}
where $\lambda $ is the wavelength. ${G^m}$ and ${G^n}$ are the antenna gains of the transmitter at $m$ and receiver at $n$ respectively. Thus, the channel gain in dB between $m$ and $n$ can be calculated, i.e., the element ${\varphi _{mn}}$ in the $m$th row and the $n$th column of matrix $\bm{\varphi}$, as
\begin{equation}
{\varphi _{mn}} =  - 10{\log _{10}}({{{p^n}} \mathord{\left/
		{\vphantom {{{p^n}} {{p^m}}}} \right.
		\kern-\nulldelimiterspace} {{p^m}}}{\text{)}}{\text{.}}
\label{eq27}
\end{equation}
where ${p^m}$ is transmitting power of the $m$th transmitting cube.

Finally, we can predict the channel gain between any two cubes in the ROI and construct a scenario-dependent dictionary matrix. In this paper, we only calculate a small part of dictionary matrix and achieve the whole matrix by interpolation. The inverse distance weighted interpolation is adopted \cite{Denkovski12CROWNCOM}. It assumes that each known value has a local influence with respect to the distance, which is consistent with the radio propagation principle. 

Let us set the data obtained by RT as ${\varphi _g},g = 1,2, \ldots ,{N_0}$, corresponding to the element in the ${g_x}$ th row and the ${g_y}$ th column of matrix $\bm{\varphi}$. The unknown value ${\varphi _{ij}}$ in the $i$th row and the $j$th column can be obtained by 
\begin{equation}
{\varphi _{ij}} = \frac{{\sum\limits_{g = 1}^{{N_0}} {\left( {d_{ij}^g} \right)_{}^{ - p}{\varphi _g}} }}{{\sum\limits_{g = 1}^{{N_0}} {\left( {d_{ij}^g} \right)_{}^{ - p}} }},
\label{eq28}
\end{equation}
where $d_{ij}^g = \sqrt {\left( {i - {g_x}} \right)_{}^2 + \left( {j - {g_y}} \right)_{}^2} $ is the distance between interpolation point and the known point, and $p$ is the distance exponent. 

\subsection{MMI-based Sampling Optimization}
The layout of sampling sensors, i.e., the measurement matrix $\bm{\psi}$, has a significant effect on the SEM construction accuracy.  In \cite{Wang04IPSL, Zhao02SPM}, the authors used the mutual information between the predicted sensors’ observation and the current target location distribution to optimize the layout.  

Different from traditional methods using the random selection, we model the sparse signal reconstruction problem as an information theory problem in communication channel, where sparse signal \bm{$\omega$} is the input to the channel and the SLs’ RSS samples \bm{$t$} is the output. The SLs are tasked to observe in order to increase the information (or to reduce the uncertainty) about the \bm{$\omega$}’s state. We select SLs based on the MMI criterion.  

We define the index set of candidate SLs for selection is ${\mathbf{S}}$. The subset of indices for the determined SLs is expressed as ${{\mathbf{S}}_k}$. When the SLs have different observation angles and perceptual uncertainties, the information gain attributable to different SLs can be quite different \cite{Saito21Access, Hintz91TSMC}. Then, the SLs set ${\mathbf{\varsigma }}$ $\left( {{\mathbf{\varsigma }} \subset {\mathbf{S}},\left| {\mathbf{\varsigma }} \right| = M} \right)$  is chosen when their observations ${\bm{t}_{\mathbf{\varsigma }}}$  minimizes the expected conditional entropy of the posterior distribution of \bm{$\omega$}, as given by
\begin{equation}
{\mathbf{\hat \varsigma }} = \arg \mathop {\min }\limits_{{\mathbf{\varsigma }} \subset {\mathbf{S}}} H\left( {\bm{\omega} |{\bm{t}_{\mathbf{\varsigma }}}} \right),
\label{eq29}
\end{equation}
which is equivalent to maximize the entropy reduction of \bm{$\omega$}. We maximize mutual information
\begin{equation}
\begin{aligned}
{\mathbf{\hat \varsigma }} &= \arg \mathop {\max }\limits_{{\mathbf{\varsigma }} \subset {\mathbf{S}}} \left\{ {H\left( \bm{\omega}  \right) - H\left( {\bm{\omega} |{\bm{t}_{\mathbf{\varsigma }}}} \right)} \right\}\\ 
&= \arg \mathop {\max }\limits_{{\mathbf{\varsigma }} \subset {\mathbf{S}}} I\left( {\bm{\omega} ;{\bm{t}_{\mathbf{\varsigma }}}} \right),
\end{aligned} 
\label{eq30}
\end{equation}
\begin{equation}
\begin{aligned}
I\left( {\bm{\omega} ;{\bm{t}_{\mathbf{\varsigma }}}} \right) &= H\left( \bm{\omega}  \right) - H\left( {\bm{\omega} |{\bm{t}_{\mathbf{\varsigma }}}} \right) \\ 
&= \frac{1}{2}\ln \left| {\mathbf{\Lambda }} \right| - \frac{1}{2}\ln \left| {\mathbf{\Sigma }} \right| \\ 
&= \frac{1}{2}\ln \left| {\frac{{\mathbf{\Lambda }}}{{{{\left( {\beta {{\mathbf{\Phi }}^{\text{T}}}{\mathbf{\Phi }} + {\mathbf{\Lambda }}} \right)}^{ - 1}}}}} \right| \\ 
\end{aligned} 
\label{eq31}
\end{equation}
The derivation of formula (31) is given in Appendex\ref{append1x a}. In the initialization stage, there is no prior information related to the sparse signal \bm{$\omega$}. According to SBL, we assign the same variance $\alpha $, i.e., 1. Then, the formula (\ref{eq31}) can be further converted to 
\begin{equation}
\begin{aligned}
I\left( {\bm{\omega} ;{\bm{t}_{\mathbf{\varsigma }}}} \right) &= \frac{1}{2}\ln \left| {\frac{{\alpha {\mathbf{I}}}}{{{{\left( {\beta {{\mathbf{\Phi }}^{\text{T}}}{\mathbf{\Phi }} + \alpha {\mathbf{I}}} \right)}^{ - 1}}}}} \right| \\ 
&= \frac{1}{2}\ln \left| {\alpha \beta {\mathbf{I}}\left( {{{\mathbf{\Phi }}^{\text{T}}}{\mathbf{\Phi }} + \alpha {\beta ^{ - 1}}{\mathbf{I}}} \right)} \right|, \\ 
\end{aligned} 
\label{eq32}
\end{equation}
with
\begin{equation}
\begin{aligned}
&\det \left( {\alpha \beta {\mathbf{I}}\left( {{{\mathbf{\Phi }}^{\text{T}}}{\mathbf{\Phi }} + \beta _{}^{ - 1}\alpha {\mathbf{I}}} \right)} \right)\\
&= {\left( {\alpha \beta } \right)^N}\det \left( {{{\mathbf{\Phi }}^{\text{T}}}{\mathbf{\Phi }} + \beta _{}^{ - 1}\alpha {\mathbf{I}}} \right) \\ 
&= {\left( \alpha  \right)^{2N}}\det \left( {\frac{\beta }{\alpha }{{\mathbf{\Phi }}^{\text{T}}}{\mathbf{\Phi }} + {\mathbf{I}}} \right) \\ 
&= {\left( \alpha  \right)^{2N}}\det \left( {\frac{\beta }{\alpha }{\mathbf{\Phi }}{{\mathbf{\Phi }}^{\text{T}}} + {\mathbf{I}}} \right) \\ 
&= {\left( \alpha  \right)^{2N - M}}{\beta ^M}\det \left( {{\mathbf{\Phi }}{{\mathbf{\Phi }}^{\text{T}}} + \frac{\alpha }{\beta }{\mathbf{I}}} \right), \\ 
\end{aligned} 
\label{eq33}
\end{equation}
where ${\raise0.7ex\hbox{$\alpha $} \!\mathord{\left/
		{\vphantom {\alpha  \beta }}\right.\kern-\nulldelimiterspace}
	\!\lower0.7ex\hbox{$\beta $}}$ is a sufficiently small number denoted the ratio of noise variance to sparse signal variance \cite{Saito21Access}. Therefore, by ignoring the constant terms, the final objective function of MMI-based sampling is asymptotically approaches to
\begin{equation}
\begin{aligned}
{\mathbf{\hat \varsigma }} &= \arg \mathop {\max }\limits_{{\mathbf{\varsigma }} \subset {\mathbf{S}}} \ln \left( {\det \left( {{\mathbf{\Phi }}{{\mathbf{\Phi }}^{\text{T}}}} \right)} \right) \\ 
&= \arg \mathop {\max }\limits_{{\mathbf{\varsigma }} \subset {\mathbf{S}}} \ln \left( {\det \left( {\bm{\psi} \bm{\varphi} {\bm{\varphi} ^{\text{T}}}{\bm{\psi} ^{\text{T}}}} \right)} \right), \\ 
\end{aligned} 
\label{eq34}
\end{equation}
where ${\mathbf{\Phi }} = \bm{\psi} \bm{\varphi}$. $\psi  = \left( {\mathbf{I}_{N} }\right)_{\mathbf{\varsigma }\bm{.}}$ is consisted of the rows indexed by set ${\mathbf{\varsigma }}$ in ${{\mathbf{I}}_N}$. Accordingly, the problem can be further expressed as
\begin{equation}
\begin{gathered}
\bm{\psi}  = \arg \mathop {\max }\limits_{{\mathbf{\varsigma }} \subset {\mathbf{S}}} \ln \left( {\det \left( {{\mathbf{\Phi }}{{\mathbf{\Phi }}^{\text{T}}}} \right)} \right), \hfill \\
{\text{s}}{\text{.t}}{\text{. }} \bm{\psi}  = \left( {\mathbf{I}_{N} }\right)_{\mathbf{\varsigma }\bm{.}}\hfill \\ 
\end{gathered} 
\label{eq35}
\end{equation}
We solve the above problem by the greedy algorithm. Let ${{\mathbf{\Phi }}_t}$ denote the sensing matrix after the $t$th SL’s selection as
\begin{equation}
{{\mathbf{\Phi }}_t} = \left[ {\bm{\varphi} _{{i_1}}^{\text{T}},\bm{\varphi} _{{i_2}}^{\text{T}}, \cdots ,\bm{\varphi} _{{i_{t - 1}}}^{\text{T}},\bm{\varphi} _{{i_t}}^{\text{T}}} \right]_{}^{\text{T}},
\label{eq36}
\end{equation}
where ${i_t}\left( {{i_t} \in {\mathbf{\varsigma }},{i_t} \in {\mathbf{S}}} \right)$ is the index of the $t$th selected SL and ${\bm{\varphi} _{{i_t}}}$ is the ${i_t}$th row vector of the sparse dictionary matrix $\bm{\varphi}$. The step-by-step maximization process is considered by the greedy method. The matrix can be expanded as
\begin{equation}
\begin{aligned}
&\det \left( {{{\mathbf{\Phi }}_t}{\mathbf{\Phi }}_t^{\text{T}}} \right)\\
 &= \det \left( {\left[ {\begin{array}{*{20}{c}}
		{{{\mathbf{\Phi }}_{t - 1}}} \\ 
		{{\bm{\psi} _t}} 
		\end{array}} \right]\left[ {\begin{array}{*{20}{c}}
		{{\mathbf{\Phi }}_{t - 1}^{\text{T}}}&{\bm{\psi} _t^{\text{T}}} 
		\end{array}} \right]} \right) \\ 
&= {\bm{\psi} _t}\left( {{\mathbf{I}} - {\mathbf{\Phi }}_{t - 1}^{\text{T}}{{\left( {{{\mathbf{\Phi }}_{t - 1}}{\mathbf{\Phi }}_{t - 1}^{\text{T}}} \right)}^{ - 1}}{{\mathbf{\Phi }}_{t - 1}}} \right)\bm{\psi} _t^{\text{T}}\\
& \times \det \left( {{{\mathbf{\Phi }}_{t - 1}}{\mathbf{\Phi }}_{t - 1}^{\text{T}}} \right). \\ 
\end{aligned} 
\label{eq37}
\end{equation}	
Therefore, the $t$th row of $\bm{\psi} $ is determined according to the following formula,
\begin{equation}
\begin{aligned}
{\bm{\psi} _t} &= \mathop {\arg \max }\limits_{{i_t} \in {\mathbf{S}}\backslash {{\mathbf{S}}_k}} \det \left( {{{\mathbf{\Phi }}_t}{\mathbf{\Phi }}_t^{\text{T}}} \right) \\ 
&= \mathop {\arg \max }\limits_{{i_t} \in {\mathbf{S}}\backslash {{\mathbf{S}}_k}} {\bm{\psi} _t}\left( {{\mathbf{I}} - {\mathbf{\Phi }}_{t - 1}^{\text{T}}{{\left( {{{\mathbf{\Phi }}_{t - 1}}{\mathbf{\Phi }}_{t - 1}^{\text{T}}} \right)}^{ - 1}}{{\mathbf{\Phi }}_{t - 1}}} \right)\bm{\psi} _t^{\text{T}}. \\ 
\end{aligned} 
\label{eq38}
\end{equation}	

After $M$ recursive iterations, the optimal $M$ SLs can be finally selected. The measurement matrix $\bm{\psi}$ is obtained.

\subsection{SEM Construction with Improved SBL}
\label{SBL}
In Section \ref{SSM}, we recover \bm{$\omega$} with the mean \bm{$\mu$} and evaluate the recovery accuracy by the variance ${\mathbf{\Sigma }}$. The hyperparameters are estimated by maximum a posterior (MAP) probability \cite{Tipping01JMLR} as
\begin{equation}
\begin{aligned}
\left( {\bm{\alpha} ,\beta } \right) &= \mathop {\arg \max }\limits_{\bm{\alpha} ,\beta } p\left( {\bm{\alpha} ,\beta |\bm{t}} \right) \\ 
&= \mathop {\arg \max }\limits_{\bm{\alpha} ,\beta } p\left( {\bm{t}|\bm{\alpha} ,\beta } \right)p\left( \bm{\alpha}  \right)p\left( \beta  \right) \\ 
&= \mathop {\arg \max }\limits_{\bm{\alpha} ,\beta } \ln p\left( {\bm{t}|\bm{\alpha} ,\beta } \right)p\left( \bm{\alpha}  \right)p\left( \beta  \right), \\ 
\end{aligned} 
\label{eq39}
\end{equation}	
which is equivalent to maximize the product of marginal likelihood $p\left( {\bm{t}|\bm{\alpha} ,\beta } \right)$ and the priors over hyperparameters in the logarithmic case. By ignoring the irrelevant terms, we can obtain the objective equation
\begin{equation}
\begin{gathered}
\mathcal{L}\left( {\bm{\alpha} ,\beta } \right) =  - \frac{1}{2}\left\{ {\log \left| {\mathbf{C}} \right| + \bm{t}_{}^T\left( {\mathbf{C}} \right)_{}^{ - 1}\bm{t}} \right\} \\
+ \sum\limits_{i = 1}^N {\left( {a\log {\alpha _i} - b{\alpha _i}} \right)}  + c\log \beta  - d\beta , \\ 
{\mathbf{C}} = \beta _{}^{ - 1}{\mathbf{I}} + {\mathbf{\Phi \Lambda }}_{}^{ - 1}{{\mathbf{\Phi }}^T}. \\ 
\end{gathered} 
\label{eq40}
\end{equation}	

We exploit expectation-maximization (EM) to solve \bm{$\alpha$} and $\beta $. For \bm{$\alpha$}, the update procedure is equivalent to maximize $E_{\bm{\omega} |{\mathbf{t}},\bm{\alpha} ,\beta }^{}\left[ {\log p\left( {\bm{\omega} |\bm{\alpha} } \right)p\left( \bm{\alpha}  \right)} \right]$, and then the update rule can be derived through differentiation
\begin{equation}
\begin{gathered}
{\alpha _i} = \frac{{1 + 2a}}{{\left\langle {\omega _i^2} \right\rangle  + 2b}}, \hfill \\
\left\langle {\omega _i^2} \right\rangle  = \mu _i^2 + {\Sigma _{i,i}}, \hfill \\ 
\end{gathered} 
\label{eq41}
\end{equation}	
where ${\Sigma _{i,i}}$ represents the $i$ th diagonal element of ${\mathbf{\Sigma }}$. We maximize $E_{\bm{\omega} |\bm{t},{\mathbf{\alpha }},\beta }^{}\left[ {\log p\left( {\bm{t}|\bm{\omega} ,\beta } \right)p\left( \beta  \right)} \right]$ to update $\beta $ as
\begin{equation}
\beta _{}^{new} = \frac{{M + 2c}}{{\left\| {\bm{t} - {\mathbf{\Phi }}\bm{\mu} } \right\|_{}^2 + \left( {\beta _{}^{old}} \right)_{}^{ - 1}\sum\limits_i {1 - \alpha _i^{}{\Sigma _{i,i}}}  + 2d}}.
\label{eq42}
\end{equation}	
The derivation of (\ref{eq41}) and (\ref{eq42}) are given in Appendix\ref{append1x b}. Continue the above iterations between (\ref{eq18}), (\ref{eq41}) and (\ref{eq42}) until the convergence condition is satisfied, and then we can obtain the solution of \bm{$\omega$} by \bm{$\mu$}.

Furthermore, in the traditional SBL algorithm, a fix threshold $thre\_\alpha _{}^{ - 1}$ is set to prune the small values of the recovered sparse signal in each iteration. The ${\mu _i}$ equals to zero when $\alpha _i^{ - 1} < thre\_\alpha _{}^{ - 1}$, and thus the corresponding $i$th column in ${\mathbf{\Phi }}$ can be pruned out. The recovery algorithm based on the traditional SBL is unable to accurately restore the locations of all sources in the 3D SEM. Besides, it is difficult to determine the value of $thre\_\alpha _{}^{ - 1}$ under different scenarios. However, the positions of the recovered signal sources are usually adjacent or close to the real signal sources in Cartesian coordinate. Therefore, we develop the clustering-based SBL (CSBL) algorithm and use adaptive threshold to improve the recovery accuracy.

Firstly, we define an adaptive dynamic threshold truncation in the pruning step of algorithm iterations, which is
\begin{equation}
thre\_\alpha _{}^{ - 1} = {\text{mean}}\left( {{{\bm{\hat \alpha} }^{ - 1}}} \right) - {\text{std}}\left( {{{\bm{\hat \alpha} }^{ - 1}}} \right).
\label{eq43}
\end{equation}	

Through the adaptive dynamic threshold truncation, the $\alpha _i^{ - 1}$ near the source points are selected to the most extent and the cube whose power is below threshold can be abandoned. The pruning rule is
\begin{equation}
{\alpha _i^{ - 1} } = \begin{cases}
{\alpha _i^{ - 1},}&{\text{if }}{\alpha _i^{ - 1}} > thre\_\alpha _{}^{ - 1}, \\ 
{0,}&{\text{otherwise.}} 
\end{cases}
\label{eq44}
\end{equation}

Then, we propose to combine clustering algorithm with the SBL. Considering the number of transmitters is unknown, we employ the MMD clustering algorithm. It can adaptively determine the cluster seeds when the number of clusters is unknown and improve the efficiency of partitioning the dataset. The MMD clustering is a trial-based clustering algorithm in pattern recognition. It takes the furthest object as the clustering center based on Euclidean distance, and can avoid the situation that the cluster seeds may be too close when the initial value is selected by k-means method. 

At the end of the SBL iteration process, the solution \bm{$\omega$} may only have a small number of significant coefficients. The rest of small coefficients, which contribute very little to the sparse signal, is negligible. Therefore, we first approximate \bm{$\omega$} by neglecting the small coefficients as
\begin{equation}
{\omega _i} = \begin{cases}
{0,}&{20{{\log }_{10}}\left( {\frac{{{\omega _i}}}{{\mathop {\max }\limits_j ({\omega _j})}}} \right) < \delta ,} \\ 
{{\omega _i},}&{\text{otherwise.}} 
\end{cases}
\label{eq45}
\end{equation}
where $\delta $ is a negative sparsity threshold. Accordingly, the $q$ non-zero item set $\Im  = \left\{ {{\zeta _1},{\zeta _2}, \ldots ,{\zeta _q}} \right\}$ of \bm{$\omega$} estimated preliminarily is obtained, which represents the cubes in 3D space. Due to the characteristics of clusters they are distributed in space, the candidate set $\Im $ will be divided into $K $ clusters denoted by ${\aleph _1},{\aleph _2}, \ldots ,{\aleph _K}$ adaptively by MMD clustering. Then, by weighting the items in $\Im $ with the averaging rule, the cluster centers are obtained as 
\begin{equation}
\begin{array}{*{20}{c}}
{\left\{ \begin{gathered}
	local_k^{{x_0}} = \frac{{\sum\limits_{t \in {\aleph _k}} {\omega _t^{} \cdot local_t^x} }}{{\sum\limits_{t \in {\aleph _k}} {\omega _t^{}} }} \hfill \\
	local_k^{{y_0}} = \frac{{\sum\limits_{t \in {\aleph _k}} {\omega _t^{} \cdot local_t^y} }}{{\sum\limits_{t \in {\aleph _k}} {\omega _t^{}} }} \hfill \\
	local_k^{{z_0}} = \frac{{\sum\limits_{t \in {\aleph _k}} {\omega _t^{} \cdot local_t^z} }}{{\sum\limits_{t \in {\aleph _k}} {\omega _t^{}} }} \hfill \\ 
	\end{gathered}  \right.,}&{k = 1,2, \ldots ,K,} 
\end{array}
\label{eq46}
\end{equation}	
where $local_t^x$, $local_t^y$ and $local_t^z$ are the $x$, $y$, $z$ coordinates of the $t$ th cube of set ${\aleph _k}$ in 3D space. $local_k^{{x_0}}$, $local_k^{{y_0}}$ and $local_k^{{z_0}}$ are the $x$, $y$, $z$ coordinates of the cluster center of set ${\aleph _k}$, which is also the re-estimated non-zero position in the updated sparse signal ${\bm{\omega} ^ * }$. Simultaneously, we update the sparse coefficient as
\begin{equation}
\omega _k^ *  = \frac{{\sum\limits_{t \in {\aleph _k}} {\omega _t^{} \cdot \omega _t^{}} }}{{\sum\limits_{t \in {\aleph _k}} {\omega _t^{}} }}
\label{eq47}
\end{equation}

Eventually, the updated sparse signal ${\bm{\omega} ^ * }$ is obtained. According to the recovered sparse signal, the SEM construction model is
\begin{equation}
{\mathbf{X}} = \bm{\varphi} {\bm{\omega} ^ * },
\label{eq48}
\end{equation}
where ${\mathbf{X}}$ is the spectrum vector we constructed.

\begin{table}[!t] 
	\caption{The main simulation parameters}
	\renewcommand\arraystretch{1.4}
	\centering
	\begin{tabular}{|c||c|}
		\hline
		Parameters                                           & Value   \\
		\hline     
		Region of interest                                    & $100{\text{m}} \times 100{\text{m}} \times 50{\text{m}}$ \\
		\hline     
		SEM tensor size                                       & $10 \times 10 \times 10$   \\
		\hline     
		Granularity of SEM tensor                             & $10{\text{m}} \times 10{\text{m}} \times 5{\text{m}}$  \\
		\hline     
		Number of RF transmitters ($K$)                       & 4, 8, 12, 16    \\
		\hline     
		Transmitting frequency ($f$)                          & 1 GHz   \\
		\hline     
		Transmitting power ($P_{}^t$)                         & 2 W    \\
		\hline     
		Positions of RF transmitters                          & Random generated in the ROI   \\
		\hline     
		\begin{tabular}[c]{@{}c@{}c@{}}Sampling rate and \\ number of RF transmitters \\ in SEM construction performance\end{tabular} 
		& \begin{tabular}[c]{@{}c@{}c@{}}  $r = 0.1$, $K = 4$ \\ \end{tabular}    \\
		\hline     
		Algorithms for comparison                             & \begin{tabular}[c]{@{}c@{}}Random-SBL, Random-CSBL,\\ Random-MSBL, MMI-SBL,\\ MMI-CMSBL, Random-SWOMP\end{tabular}   \\
		\hline     
	\end{tabular}
\end{table}

\section{Simulation Results And Validations}
\subsection{Experiment Setup}
In this section, the performance of proposed 3D SEM construction method is analyzed and verified by simulations under the campus scenario, as shown in Fig. 5. The ROI is $100{\text{ m}} \times 100{\text{ m}} \times 50{\text{ m}}$. We firstly discretize the ROI into $10 \times 10 \times 10$ cubes and each cube is $10{\text{ m}} \times 10{\text{ m}} \times 5{\text{ m}}$. Then we construct a spectrum tensor $\bm{\chi}  \in {\Re ^{10 \times 10 \times 10}}$. We denote the proposed algorithm as MMI-CMSBL. Six SEM construction algorithms, i.e., Random-SBL, Random-CSBL, Random-MSBL, MMI-SBL, MMI-CMSBL and random stagewise weak orthogonal matching pursuit (Random-SWOMP) \cite{Blumensath09TSP} are used to construct the 3D SEM. The main simulation parameters are shown in the Table 1, where the random means that SLs are selected randomly from the ROI.

\subsection{Sparse Signal Recovery Performance}
We compare the proposed MMI-CMSBL with Random-SBL, Random-CSBL, Random-MSBL, MMI-SBL, and Random-SWOMP in terms of the SSR performance. The Mean Squared Error (MSE) of sparse signal recovery is defined as
\begin{equation}
d\left( {{\bm{\omega } ^{{\text{est}}}},{\bm{\omega } ^{{\text{true}}}}} \right) = 10\log _{10}^{}\left( {{\raise0.7ex\hbox{${\left\| {{\bm{\omega } ^{{\text{est}}}} - {\bm{\omega } ^{{\text{true}}}}} \right\|_2^{}}$} \!\mathord{\left/
			{\vphantom {{\left\| {{\bm{\omega } ^{{\text{est}}}} - {\bm{\omega } ^{{\text{true}}}}} \right\|_2^{}} {\left\| {{\bm{\omega } ^{{\text{true}}}}} \right\|_2^{}}}}\right.\kern-\nulldelimiterspace}
		\!\lower0.7ex\hbox{${\left\| {{\bm{\omega } ^{{\text{true}}}}} \right\|_2^{}}$}}} \right),
\label{eq49}
\end{equation}
where ${\bm{\omega } ^{{\text{est}}}}$ and ${\bm{\omega } ^{{\text{true}}}}$ are the estimated sparse signal and true sparse signal. 

It can be seen from the simulation results in the Fig. \ref{fig6} that the MSE of sparse signal recovery decreases as the sampling rates increases. The proposed MMI-CSBL outperforms other algorithms in terms of the convergence speed. The performance of Random-MSBL algorithm is better than that of the Random-CSBL, MMI-SBL and Random-SBL, which reveals that the propagation model-based SBL algorithm brings more performance improvement than traditional SBL in SSR by considering scenario to construct the sparse dictionary. Besides, the SBL has better performance than other CS algorithms when recovering sparse signals if the sensing matrix has high correlation.

\begin{figure}[!t]
	\centering
	\includegraphics[width=3in]{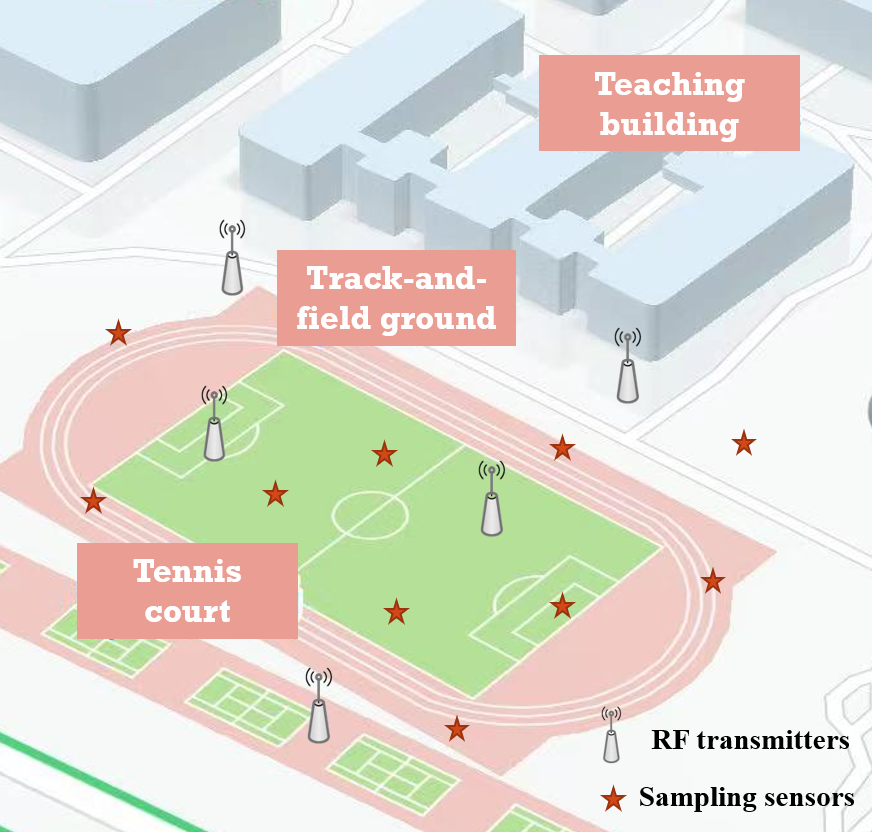}
	\caption{3D SEM construction scenario.}
	\label{fig5}
\end{figure}

\begin{figure}[!t]
	\centering
	\includegraphics[width=3.5in]{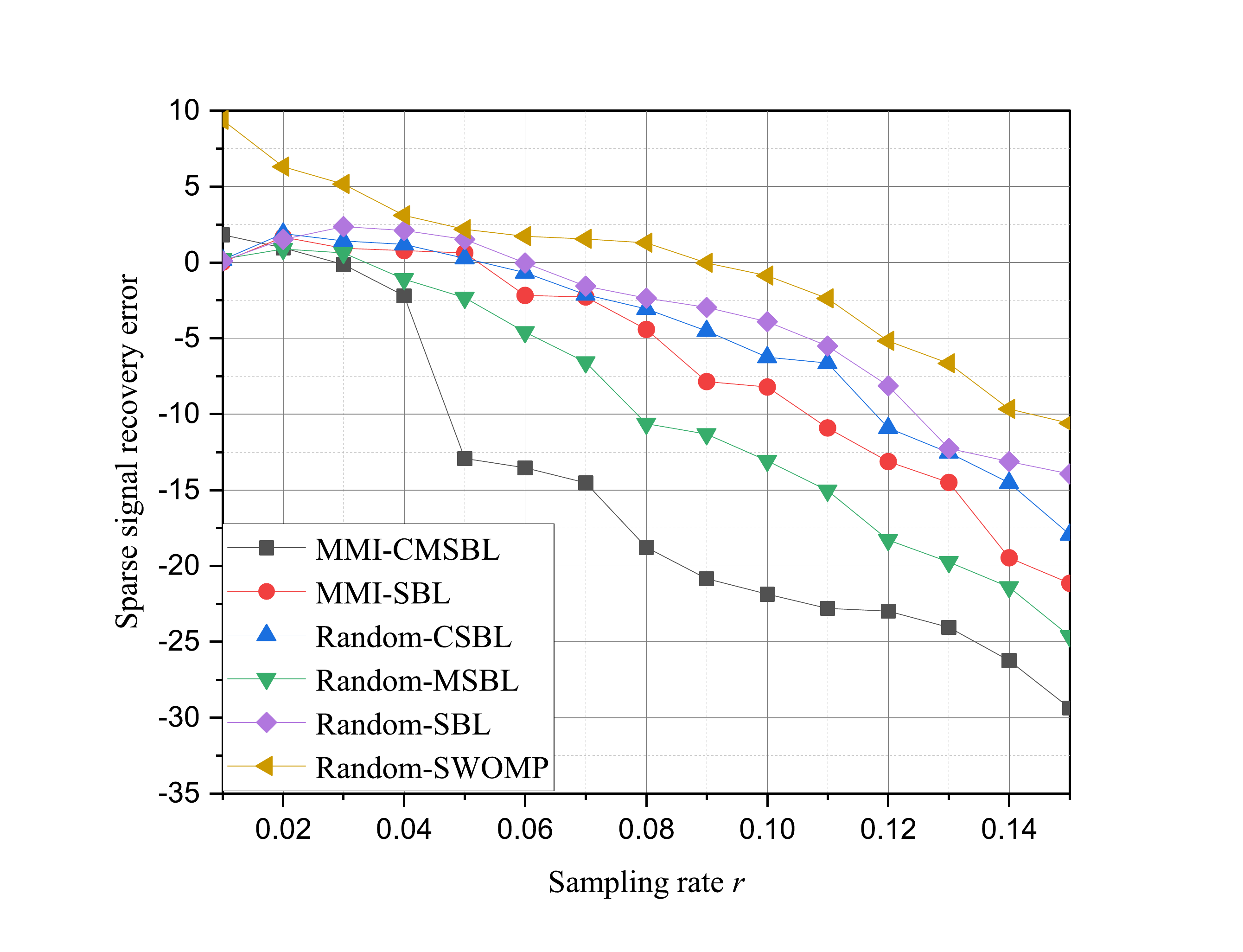}
	\caption{The MSE of sparse signal recovery performance comparisons ($K = 4$).}
	\label{fig6}
\end{figure}

\begin{figure*} 
	\centering
	\includegraphics[width=7in]{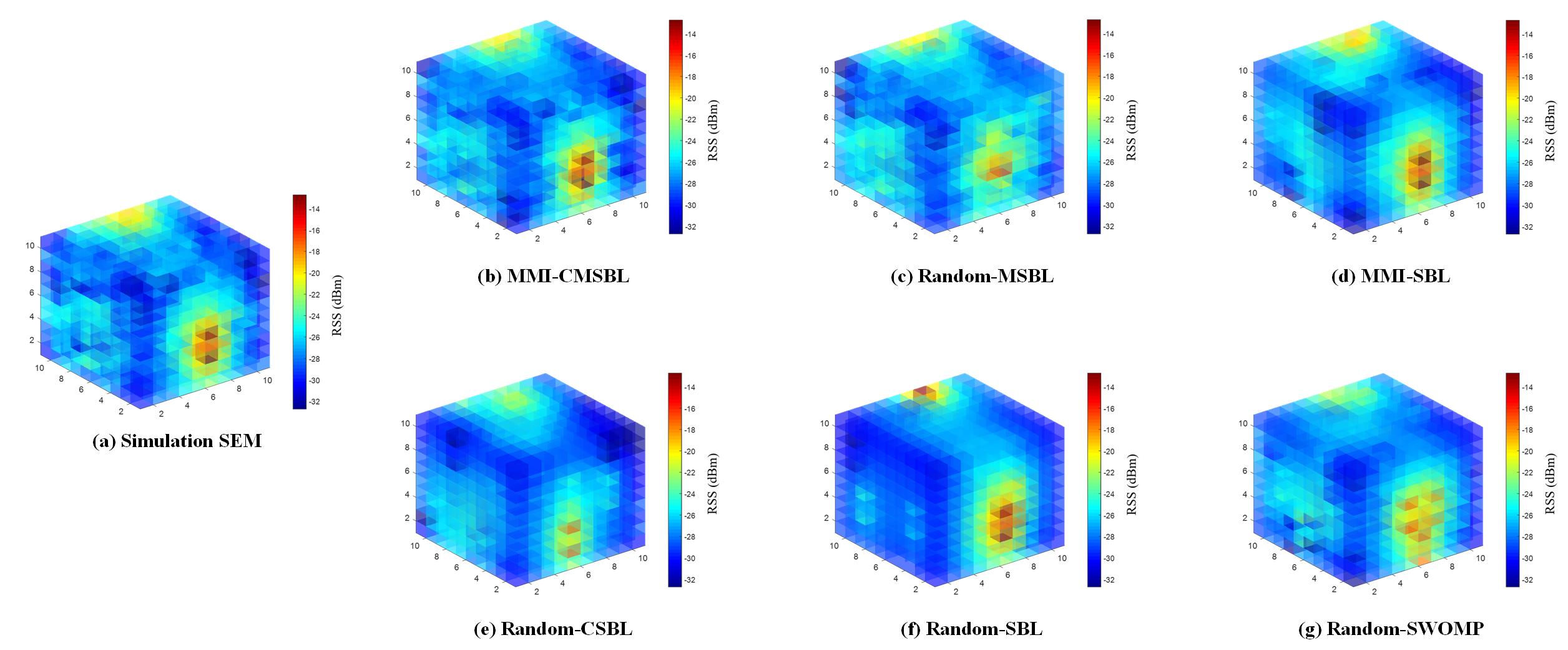}
	\caption{3D SEM construction performance visualization ($K = 4$, $r = 0.1$). (a) The true SEM; and (b) - (g) The spectrum situation recovery of MMI-CMSBL, Random-MSBL, MMI-SBL, Random-CSBL, Random-SBL and Random-SWOMP, respectively.}	
	\label{fig7}	
\end{figure*}

\subsection{SEM Construction Performance}
Fig. \ref{fig7} presents the SEM construction results of Random-SBL, Random-CSBL, Random-MSBL, MMI-SBL, MMI-CMSBL and Random-SWOMP. We can also see that the sparse signal recovery accuracy of MMI-CMSBL is higher than other algorithms, and the minimal components in Fig. \ref{fig7} (b) are reduced obviously through clustering and dynamic threshold operations.

\begin{figure}[!t]
	\centering
	\includegraphics[width=3.5in]{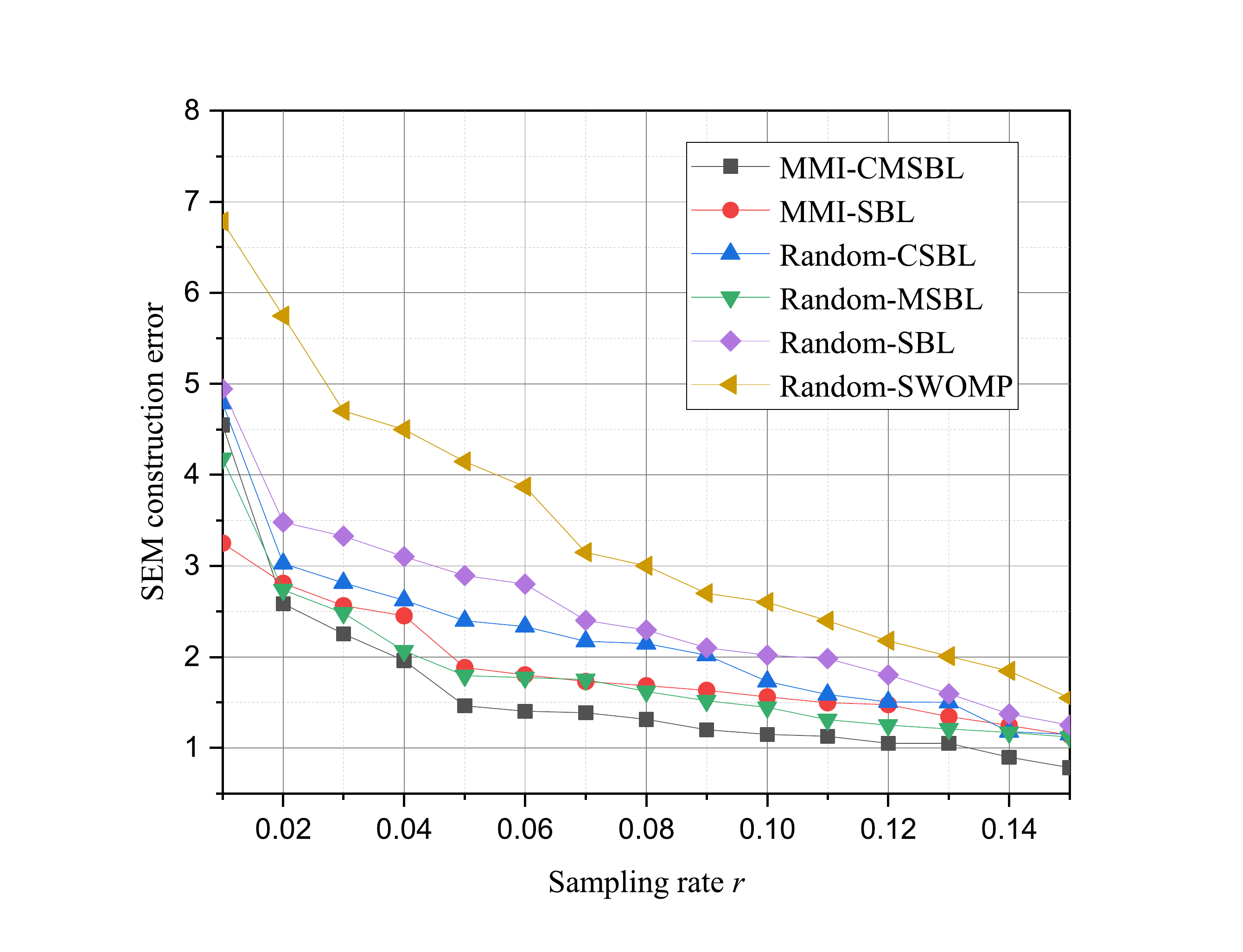}
	\caption{3D SEM construction performance comparisons ($K = 4$).}
	\label{fig8}
\end{figure}

The Root Mean Square Error (RMSE) is used to evaluate the RSS recovery performance. The RMSE is defined as the difference of each cube in RSS between the estimated REM and the real REM, given by
\begin{equation}
{\text{RMSE = }}\sqrt {\frac{1}{N}\sum\limits_{i = 1}^N {{{\left| {P{{_i^r}^{{\text{est}}}} - P{{_i^r}^{{\text{true}}}}} \right|}^2}} } ,
\label{eq50}
\end{equation}
where ${P{_i^r}}^{{\text{est}}}$ and ${P{_i^r}}^{{\text{true}}}$ are the estimated and the actual RSS values at the  th cube respectively. It can be seen from Fig. 8 that MMI-CMSBL achieves the best performance compared with other methods even at a very low sampling rate. SWOMP performs worse than other SBL-based algorithms, due to the highly correlation of the sparse dictionary. Furthermore, based on the sparse dictionary constructed by RT technology, the MMI-CMSBL and Random-MSBL algorithm can rapidly converge and accurately recover the spectrum map at a low sampling rate. Comparing Fig. \ref{fig6} and Fig. \ref{fig8}, we can see that when the MSE of sparse signal is large, the SEM construction performance is unsatisfactory. 

\subsection{Impact of Sparsity}
\begin{figure}[!t]
	\centering
	\includegraphics[width=3.5in]{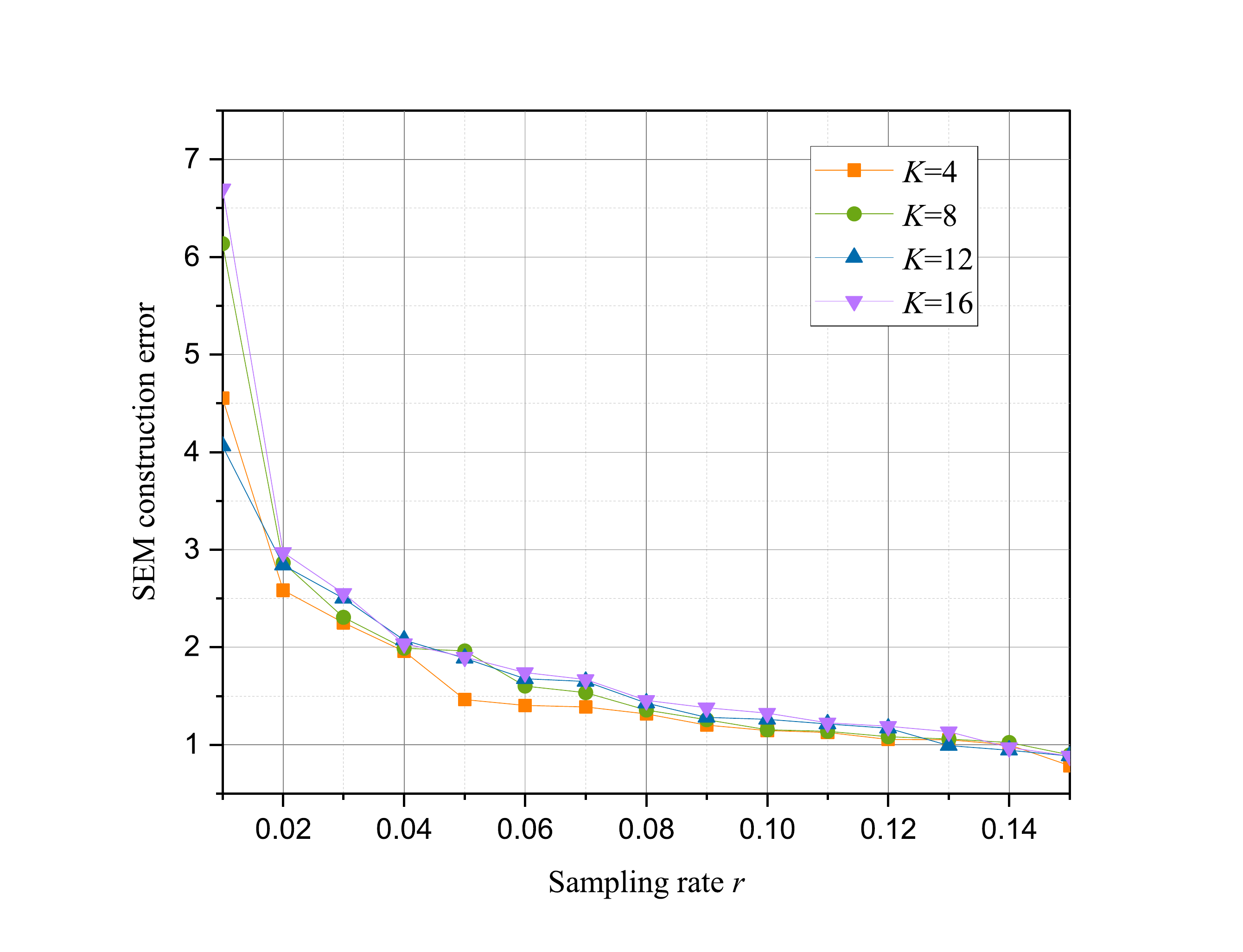}
	\caption{Impact of sparsity $K$ on the MMI-CMSBL performances of SEM construction.}
	\label{fig9}
\end{figure}

\begin{figure}[!t]
	\centering
	\includegraphics[width=3.5in]{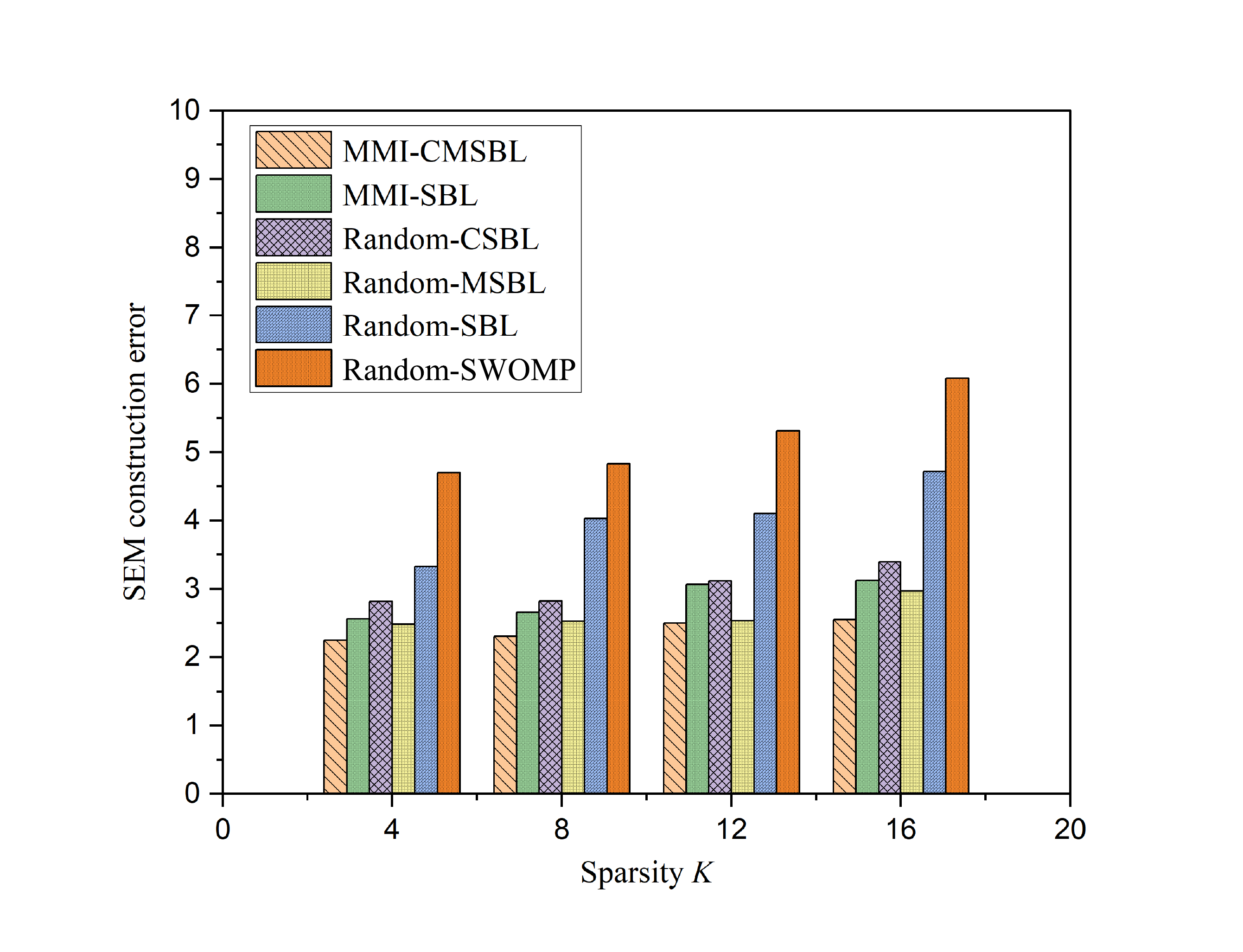}
	\caption{Impact of sparsity $K$ on the SEM construction performances of different algorithms ($r = 0.03$).}
	\label{fig10}
\end{figure}

\begin{figure}[!t]
	\centering
	\vspace{-0.5cm}
	\includegraphics[width=3.5in]{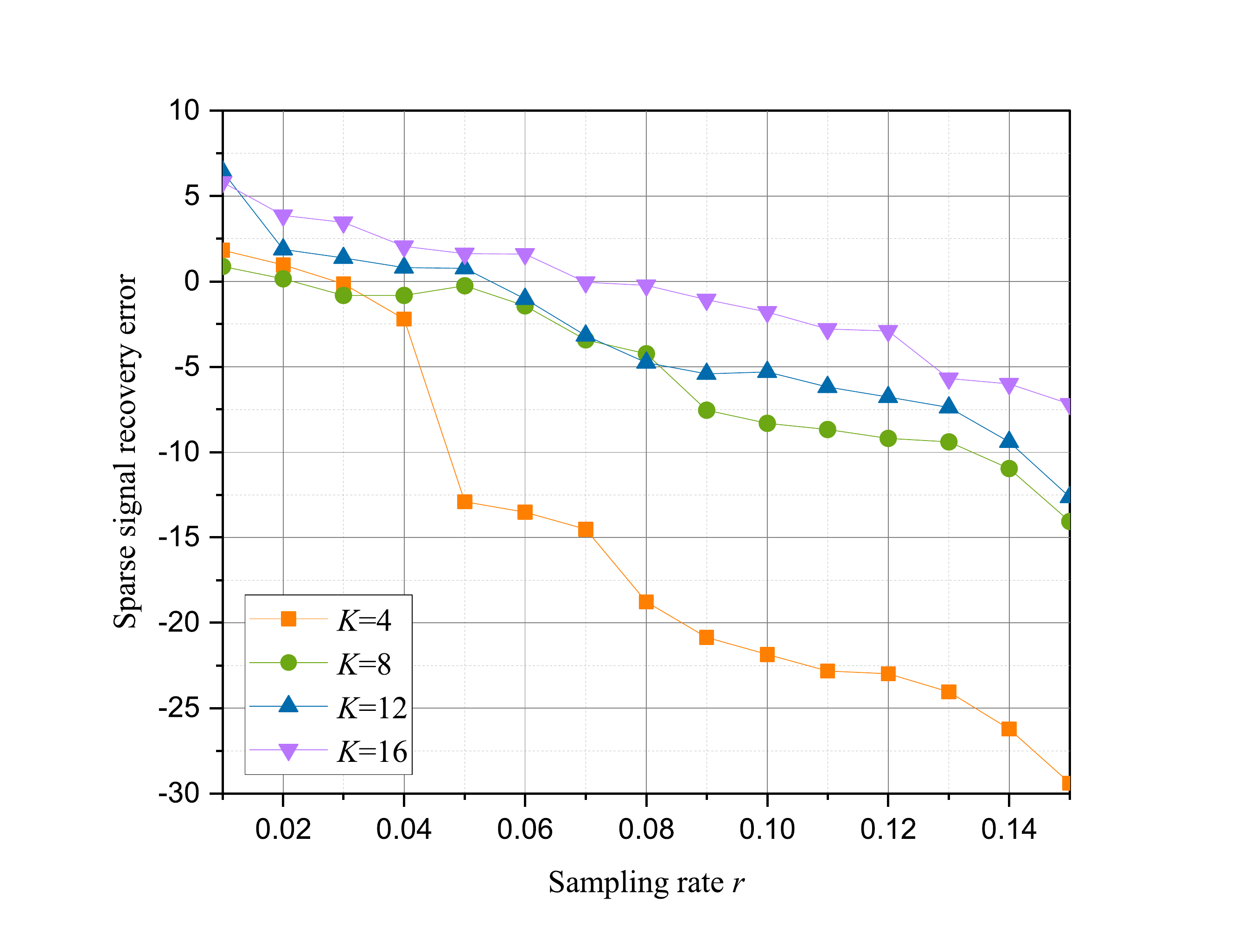}
	\caption{Impact of sparsity $K$ on the MMI-CMSBL performances of sparse signal recovery.}
	\label{fig11}
\end{figure}

\begin{figure}[!t]
	\centering
	\includegraphics[width=3.5in]{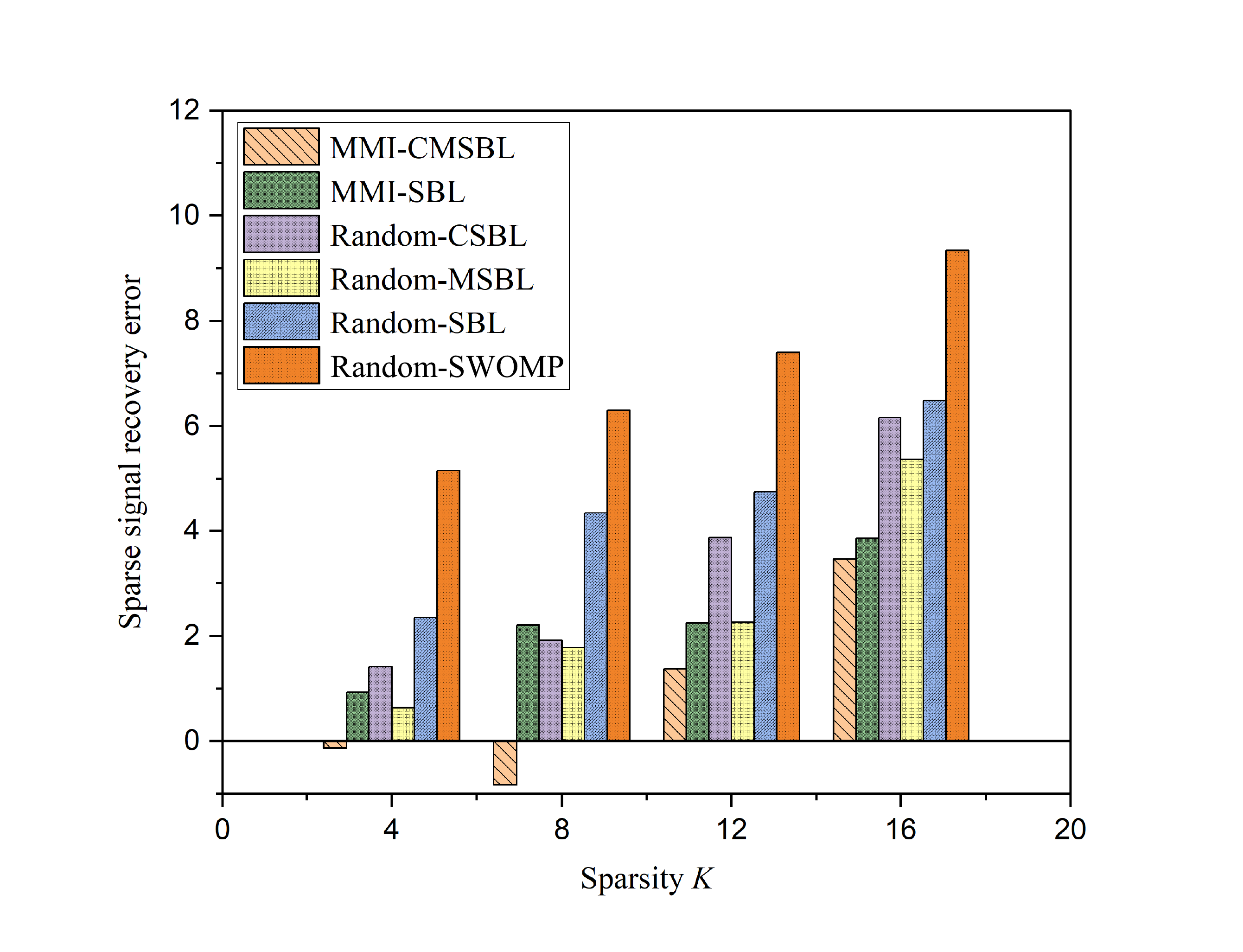}
	\caption{Impact of sparsity $K$ on the sparse signal recovery of different algorithms ($r = 0.03$).}
	\label{fig12}
\end{figure}

The impact of $K$ on the performance of SEM construction and sparse signal recovery are shown in Figs. \ref{fig9} - \ref{fig12}. In Fig. \ref{fig9} and \ref{fig11} show that as the sparsity $K$ increases, the SEM construction error and the MSE of sparse signal recovery with MMI-CMSBL increase. Meanwhile, as shown in Fig. \ref{fig9}, when sampling rate is small, the SEM construction error first increases and then decreases with the increasing of $K$. Fig. \ref{fig11} shows that the MSE of sparse signal recovery increases as $K$ increases, and we can successfully recover the signal at least when $M > 2K\ln \left( {N/K} \right)$. In addition, the convergence rate of sparse support distortion improves with the decrease of sparsity $K$. 

\begin{figure}[!t]
	\centering
	\includegraphics[width=3.5in]{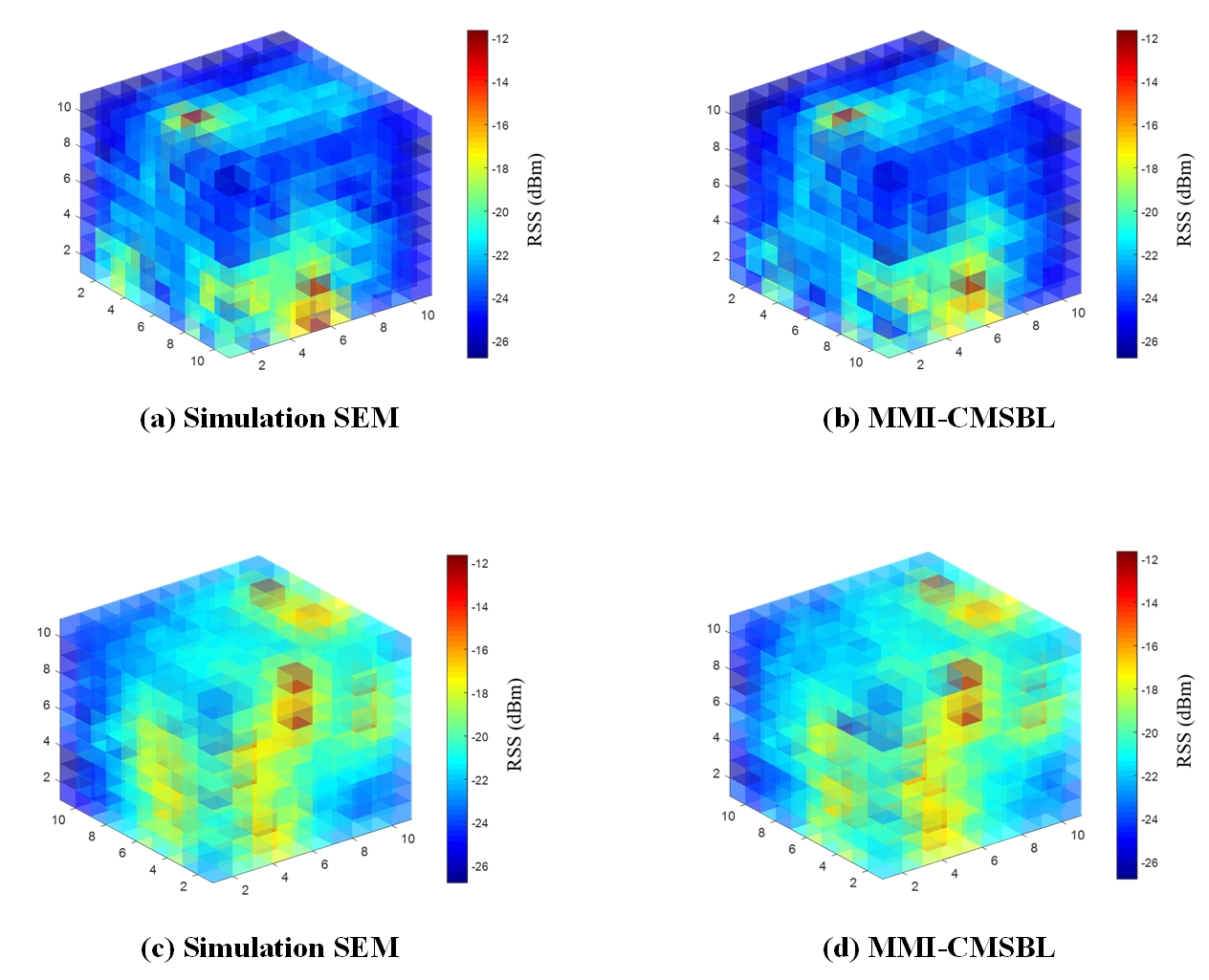}
	\caption{3D SEM construction performance visualization with different sparsity $K$ ($r = 0.1$). (a) and (c) are the simulation SEM with $K = 8$ and $K = 12$, respectively. (b) and (d) present the SEM construction of MMI-CMSBL with $K = 8$ and $K = 12$, respectively.}
	\label{fig13}
\end{figure}

In Fig. \ref{fig10} and \ref{fig12}, it can be seen that with the increase of sparsity $K$ the MMI-CMSBL can still maintain excellent performance on SEM recovery and sparse signal recovery compared with other algorithms at a fixed low sampling rate. MMI-CMSBL and Random-MSBL are less influenced by the sparsity $K$ than Random-SBL. Due to the increase of RF transmitters, their interference with each other also increases. Additionally, CSBL can achieve better performance than the traditional SBL. The visualization of SEM construction with different sparsity is shown in Fig. \ref{fig13}.

\section{Conclusion}
In this paper, we have investigated the issue of 3D SEM construction based on SBL, which offers high value for efficient application of CR. We have first formulated 3D SEM construction as a sparse sampling problem by exploiting the underlying sparse nature of 3D spectrum situation. Then, we have proposed a 3D scenario-dependent SEM construction scheme, which is composed of three components: sparse dictionary construction, sampling optimization and spectrum situation recovery. Firstly, considering the complexity of electromagnetic propagation environment in the actual scenario, we have designed a scenario-dependent sparse dictionary for SEM construction based on the channel model. Secondly, we have developed MMI-based sampling architecture to obtain optimized measurement matrix. Based on SBL framework, we have derived the optimization function of MMI sampling and have solved it by greedy algorithm. Finally, due to the ineffectiveness of traditional SBL recovery algorithm in 3D SEM construction, a tailored MMD-clustering based SBL algorithm has been proposed. The sparse signal recovered can achieve high precision by dynamic threshold pruning. We have also compared the sparse signal recovery performance and SEM construction performance among six methods, i.e., Random-SBL, Random-CSBL, Random-MSBL, MMI-SBL, MMI-CMSBL and Random-SWOMP. The impact of sparsity   on situation recovery precision has been studied. Simulations have demonstrated the superiority of the proposed 3D SEM construction scheme.

\appendices
\section{Proof of the equation (\ref{eq31})}
\label{append1x a}
In this part, we present the derivation of (\ref{eq31}) in the text. Firstly, we introduce the Mahalanobis transformation lemma,

\emph{Lemma 1.} An arbitrary $N$ dimension Gaussian distribution $\vartheta  \sim \mathcal{N}\left( {\mu ,\Sigma } \right)$, we call $\rho  = {\Sigma ^{ - \frac{1}{2}}}\left[ {\vartheta  - \mu } \right]$ the Mahalanobis transformation, where
\begin{equation}
\rho  \sim \mathcal{N}\left( {0,{I_k}} \right),
\label{eq51}
\end{equation}
which means that ${\rho _i}$ is the standard Gaussian distribution $\mathcal{N}\left( {0,1} \right)$.

\emph{Proof:} We have 
\begin{equation}
p\left( \vartheta  \right) = {\left( {2\pi } \right)^{ - \frac{N}{2}}}{\left| \Sigma  \right|^{ - \frac{1}{2}}}\exp \left[ { - \frac{1}{2}{{\left( {\vartheta  - \mu } \right)}^T}{\Sigma ^{ - 1}}\left( {\vartheta  - \mu } \right)} \right],
\label{eq52}
\end{equation}
\begin{equation}
\vartheta  = {\Sigma ^{\frac{1}{2}}}\rho  + \mu ,
\label{eq53}
\end{equation}
we compute the Jacobian determinant
\begin{equation}
{\mathbf{J}} = \det \left[ {\frac{{\partial \vartheta }}{{\partial {\rho ^T}}}} \right] = {\left| \Sigma  \right|^{\frac{1}{2}}}.
\label{eq54}
\end{equation}

According to Lemma of substitution of variables
\begin{equation}
f\left( \rho  \right) = {f_\vartheta }\left[ {h\left( \rho  \right)} \right] \times \left| {\mathbf{J}} \right|,
\label{eq55}
\end{equation}
where $\vartheta  \sim f\left( \vartheta  \right)$ and inverse function $\vartheta  = h\left( \rho  \right)$. $\left| {\mathbf{J}} \right|$ is the absolute value of the Jacobian determinant. Then the distribution of $\rho $ is 
\begin{equation}
p\left( \rho  \right) = {\left( {2\pi } \right)^{ - \frac{N}{2}}}\exp \left[ { - \frac{1}{2}{\rho ^T}\rho } \right].
\label{eq56}
\end{equation}
The lemma has been proven. Then, for continuous random variables \bm{$\omega$}, we have its conditional entropy
\begin{equation}
H\left( {\bm{\omega} |{\mathbf{t}}} \right) =  - \int {p\left( \bm{t} \right)} \int {p\left( {\bm{\omega} |\bm{t}} \right)\ln p\left( {\bm{\omega} |\bm{t}} \right)} d\bm{\omega} d\bm{t}.
\label{eq57}
\end{equation}
The posterior distribution of \bm{$\omega$} obeys the multidimensional Gaussian distribution $\mathcal{N}(\bm{\mu} ,{\mathbf{\Sigma }})$ with mean and covariance given by (\ref{eq18}). Therefore, we use a posterior distribution to approximate the distribution of \bm{$\omega$}, as SBL use a posterior mean to estimate \bm{$\omega$}, so we have
\begin{equation}
\begin{aligned}
H\left( {\bm{\omega} |\bm{t}} \right) &= {H_{\bm{t}}}\left( \bm{\omega}  \right) \\ 
&=  - \int {p\left( \bm{\omega}  \right)} \ln ({\left( {2\pi } \right)^{ - \frac{N}{2}}}{\left| {\mathbf{\Sigma }} \right|^{ - \frac{1}{2}}} \\ 
&\exp \left[ { - \frac{1}{2}{{\left( {\bm{\omega}  - \bm{\mu} } \right)}^T}{{\mathbf{\Sigma }}^{ - 1}}\left( {\bm{\omega}  - \bm{\mu} } \right)} \right])d\bm{\omega}  \\ 
&= \int {p\left( \bm{\omega}  \right)} [\ln \left( {{{\left( {2\pi } \right)}^{ - \frac{N}{2}}}{{\left| {\mathbf{\Sigma }} \right|}^{ - \frac{1}{2}}}} \right) \\ 
&- \frac{1}{2}{\left( {\bm{\omega}  - \bm{\mu} } \right)^T}{{\mathbf{\Sigma }}^{ - 1}}\left( {\bm{\omega}  - \bm{\mu} } \right)]d\bm{\omega}  \\ 
&= \ln \left( {{{\left( {2\pi } \right)}^{\frac{N}{2}}}{{\left| {\mathbf{\Sigma }} \right|}^{\frac{1}{2}}}} \right) \\ 
&+ \frac{1}{2}\int {p\left( \bm{\omega}  \right)\left[ {{{\left( {\bm{\omega}  - \bm{\mu} } \right)}^T}{{\mathbf{\Sigma }}^{ - 1}}\left( {\bm{\omega}  - \bm{\mu} } \right)} \right]d\bm{\omega} }  \\ 
\end{aligned}
\label{eq58}
\end{equation}
By using Lemma 1, we then have
\begin{equation}
\begin{aligned}
&\int {p\left( \bm{\omega}  \right)\left[ {{{\left( {\bm{\omega}  - \bm{\mu} } \right)}^T}{{\mathbf{\Sigma }}^{ - 1}}\left( {\bm{\omega}  - \bm{\mu} } \right)} \right]d\bm{\omega} }\\  
&= \int {p\left( {\mathbf{\nu }} \right) \times {{\mathbf{\nu }}^T}{\mathbf{\nu }}d\nu }  \\
&= \sum\limits_{i = 1}^N {{\text{E}}\left[ {\nu _i^2} \right]}  = \frac{N}{2}. 
\end{aligned}
\label{eq59}
\end{equation}
Accordingly, we can obtain the conditional entropy term in the formula (\ref{eq31}), as given by
\begin{equation}
\begin{aligned}
H\left( {\bm{\omega} |\bm{t}} \right)  &= \ln \left( {{{\left( {2\pi } \right)}^{\frac{N}{2}}}{{\left| {\mathbf{\Sigma }} \right|}^{\frac{1}{2}}}} \right) + \frac{N}{2}  \\
&= {\text{ln}}\left( {{{\left( {2\pi e} \right)}^{\frac{N}{2}}}{{\left| {\mathbf{\Sigma }} \right|}^{\frac{1}{2}}}} \right) \\
&= \frac{N}{2}\left( {\ln 2\pi  + 1} \right) + \frac{1}{2}\ln \left| {\mathbf{\Sigma }} \right|. 
\end{aligned}
\label{eq60}
\end{equation}
The derivation of $H\left( \bm{\omega}  \right)$ can be obtained by analogy.

\section{Proofs of the equations (\ref{eq41}) and (\ref{eq42})}
\label{append1x b}
To obtain Eq. (\ref{eq41}), we first define
\begin{equation}
\begin{aligned}
&\mathcal{L}\left( \bm{\alpha}  \right)  = {\text{E}}_{\bm{\omega} |\bm{t},\bm{\alpha} ,\beta }^{}\left[ {\ln p\left( {\bm{\omega} |\bm{\alpha} } \right)p\left( \bm{\alpha}  \right)} \right]  \\
&= {\text{E}}\left[{ - \frac{1}{2}\left( {\log \left| {{\mathbf{\Lambda }}_{}^{ - 1}} \right| + t_{}^T{\mathbf{\Lambda }}t} \right) + \sum\limits_{n = 0}^N {\left( {a\ln {\alpha _n} - b{\alpha _n}} \right)} } \right]\\
&+ con \\
&=  - \frac{1}{2}\sum\limits_{i = 1}^N {\left( {\ln \alpha _i^{ - 1} + {\alpha _i}\left( {\mu _i^2 + {\Sigma _{i,i}}} \right)} \right)}\\
 &+ \sum\limits_{i = 1}^N {\left( {a\ln {\alpha _i} - b{\alpha _i}} \right)}  + con, 
\end{aligned}
\label{eq61}
\end{equation}
with $con$ being the item constant to $\bm{\alpha} $. Then, we let $\partial \mathcal{L}\left( \bm{\alpha}  \right)/\partial \alpha _i^{} = 0$ to find the stationary point of $\alpha _i^{}$, as
\begin{equation}
\begin{aligned}
\partial \mathcal{L}\left( \bm{\alpha}  \right)/\partial \alpha _i^{} &= \left( {\frac{1}{2}\alpha _i^{ - 1} - \frac{1}{2}\left( {\mu _i^2 + {\Sigma _{i,i}}} \right)} \right) + \left( {a\alpha _i^{ - 1} - b} \right)\\
 &= 0.
\end{aligned}
\label{eq62}
\end{equation}
And we obtain
\begin{equation}
{\alpha _i} = \frac{{1 + 2a}}{{\mu _i^2 + {\Sigma _{i,i}} + 2b}}.
\label{eq63}
\end{equation}

To obtain Eq. (\ref{eq42}), we define
\begin{equation}
\begin{aligned}
&\mathcal{L}\left( \beta  \right)  = {\text{E}}_{\bm{\omega} |\bm{t},\bm{\alpha} ,\beta }^{}\left[ {\ln p\left( {\bm{t}|\bm{\omega} ,\beta } \right)p\left( \beta  \right)} \right]  \\
&= {\text{E}}\left[ {\frac{1}{2}\left( {\ln \left| {\mathbf{B}} \right| - \beta \left( {\bm{t} - {\mathbf{\Phi }}\bm{\omega} } \right)_{}^T\left( {\bm{t} - {\mathbf{\Phi }}\bm{\omega} } \right)} \right) + c\ln \beta  - d\beta } \right] \\
& + con \\
&= \frac{1}{2}\left( {M\ln \beta  - \beta \sum\limits_{j = 1}^M {{\text{E}}\left[ {\left( {\bm{t} - {\mathbf{\Phi }}\bm{\omega} } \right)_j^2} \right]} } \right)\\
& + c\ln \beta  - d\beta  + con. 
\end{aligned}
\label{eq64}
\end{equation}
Then, we let $\partial \mathcal{L}\left( \beta  \right)/\partial \beta  = 0$ to find the stationary point of $\beta $, and we obtain
\begin{equation}
\beta _{}^{new} = \frac{{M + 2c}}{{\sum\limits_{j = 1}^M {{\text{E}}\left[ {\left( {\bm{t} - {\mathbf{\Phi }}\bm{\omega} } \right)_j^2} \right]}  + 2d}},
\label{eq65}
\end{equation}
where
\begin{equation}
\begin{aligned}
&\sum\limits_{j = 1}^M {{\text{E}}\left[ {\left( {\bm{t} - {\mathbf{\Phi }}\bm{\omega} } \right)_j^2} \right]}\\  
&= {\text{E}}\left[ {\left\| {\bm{t} - {\mathbf{\Phi }}\bm{\omega} } \right\|_2^2} \right]  \\
&= \left\| \bm{t} \right\|_2^2 - 2\bm{t}_{}^T{\mathbf{\Phi }}\bm{\mu}  + \left\| {{\mathbf{\Phi }}\bm{\mu} } \right\|_2^2 + tr\left( {{\mathbf{\Sigma \Phi }}_{}^T{\mathbf{\Phi }}} \right) \\
&= \left\| {\bm{t} - {\mathbf{\Phi }}\bm{\mu} } \right\|_2^2 + \beta _{}^{ - 1}tr\left( {{\mathbf{I}} - {\mathbf{\Sigma \Lambda }}} \right)\\
&= \left\| {\bm{t} - {\mathbf{\Phi }}\bm{\mu} } \right\|_2^2 + \beta _{}^{ - 1}\sum\limits_{i = 1}^N {\left( {1 - \alpha _i^{}\Sigma _{ii}^{}} \right)} .
\end{aligned}
\label{eq66}
\end{equation}

%

\bibliographystyle{IEEEtran}
\bibliography{myref}


\newpage

\begin{IEEEbiography}[{\includegraphics[width=1in,height=1.25in,clip,keepaspectratio]{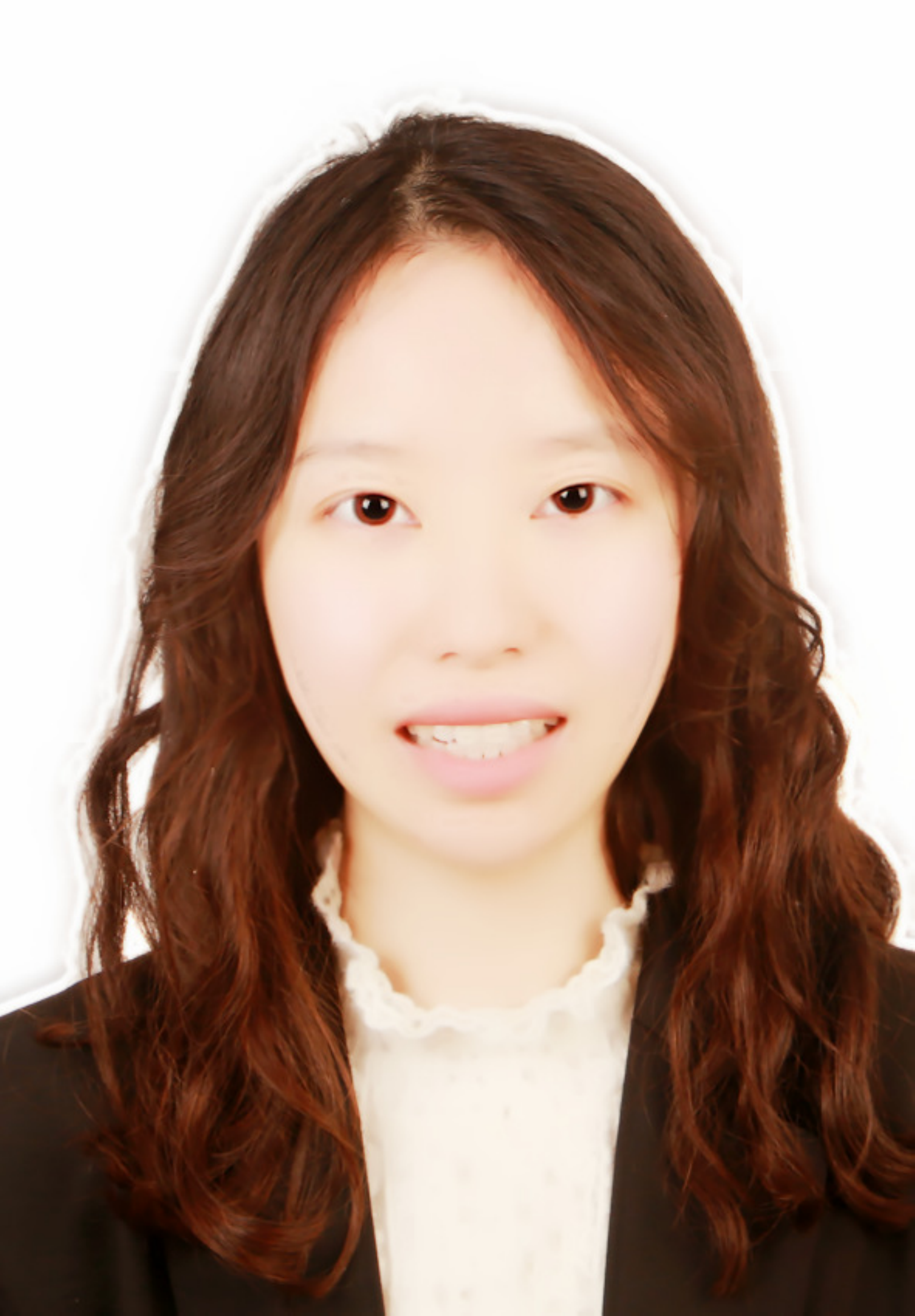}}]{Jie Wang}
received the B.S. degree in internet of things engineering from the College of Information Science and Technology, Nanjing Forestry University of China, Nanjing, China, in 2021. She is currently pursuing the Ph.D. degree in communications and information systems with the College of Electronic and Information Engineering, Nanjing University of Aeronautics and Astronautics. Her current research interests conclude spectrum mapping. 
\end{IEEEbiography}

\vspace{11pt}

\begin{IEEEbiography}[{\includegraphics[width=1in,height=1.25in,clip,keepaspectratio]{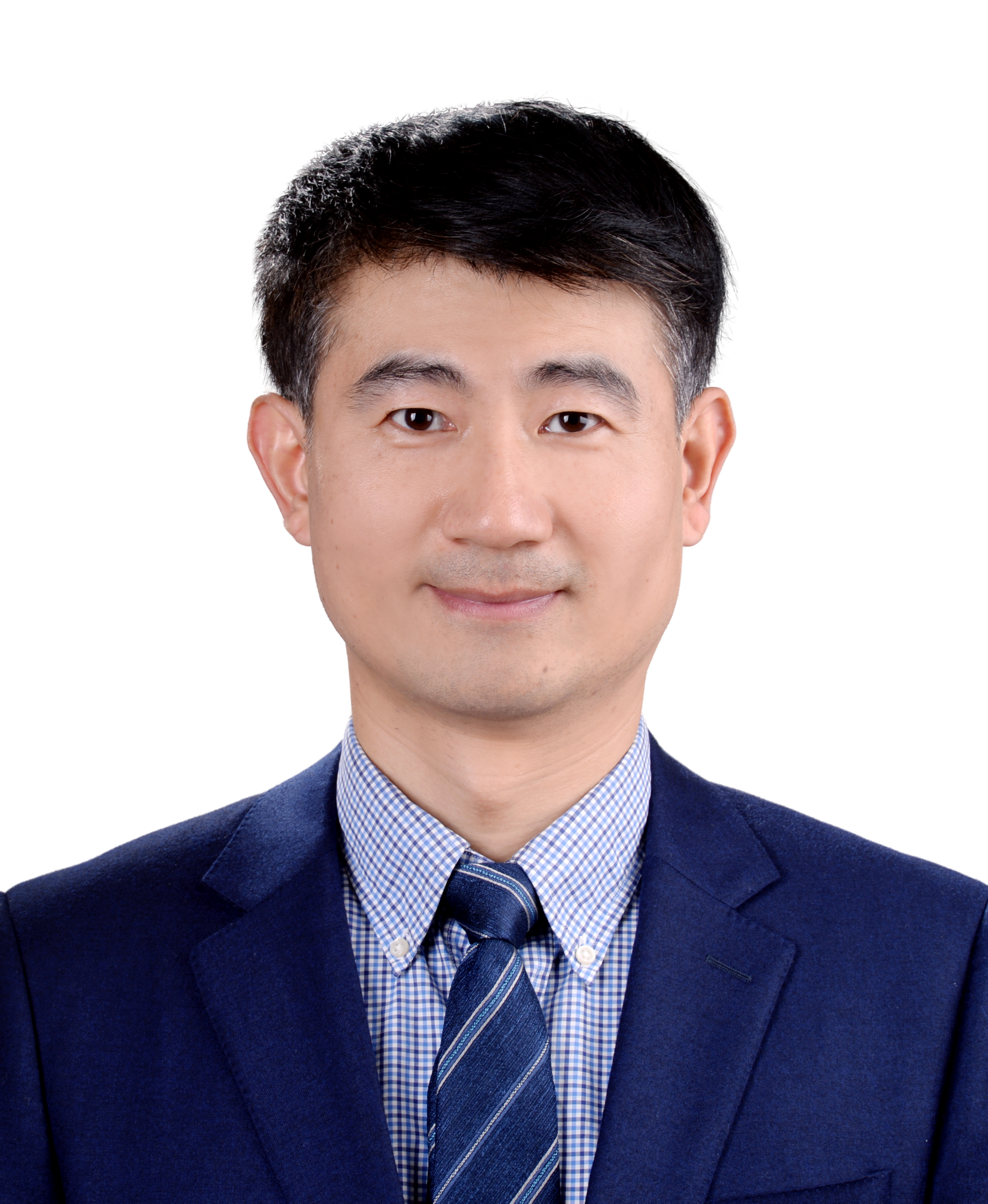}}]{Qiuming Zhu}
	received the B.S. degree in electronic engineering and the M.S. and Ph.D. degrees in communication and information system from the Nanjing University of Aeronautics and Astronautics (NUAA), Nanjing, China, in 2002, 2005, and 2012, respectively. He is currently a Professor in the College of Electronic and Information Engineering, Nanjing University of Aeronautics and Astronautics, Nanjing, China. His current research interests include channel sounding, modeling, and emulation for the fifth/sixth generation (5G/6G) mobile communication, vehicle-to-vehicle (V2V) communication, and unmanned aerial vehicles (UAV) communication systems.
\end{IEEEbiography}

\vspace{11pt}

\begin{IEEEbiography}[{\includegraphics[width=1in,height=1.25in,clip,keepaspectratio]{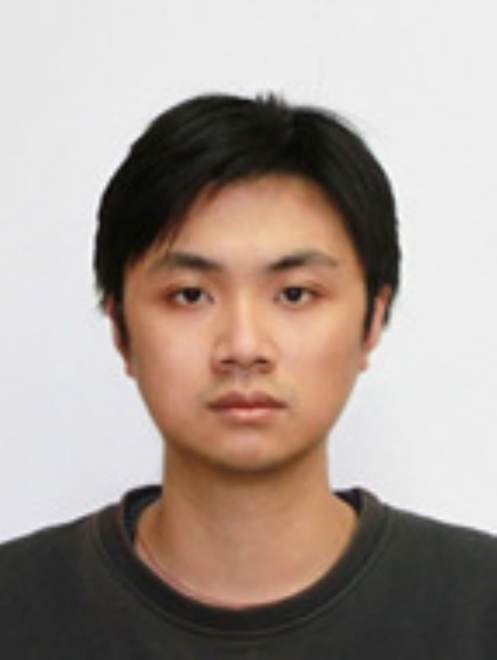}}]{Zhipeng Lin}
received the Ph.D. degrees from the School of Information and Communication Engineering, Beijing University of Posts and Telecommunications, Beijing, China, and the School of Electrical and Data Engineering, University of Technology of Sydney, NSW, Australia, in 2021. Currently, He is an Associate Researcher in the College of Electronic and Information Engineering, Nanjing University of Aeronautics and Astronautics, Nanjing, China. His current research interests include signal processing, massive MIMO, spectrum sensing, and UAV communications.
\end{IEEEbiography}

\vspace{11pt}

\begin{IEEEbiography}[{\includegraphics[width=1in,height=1.25in,clip,keepaspectratio]{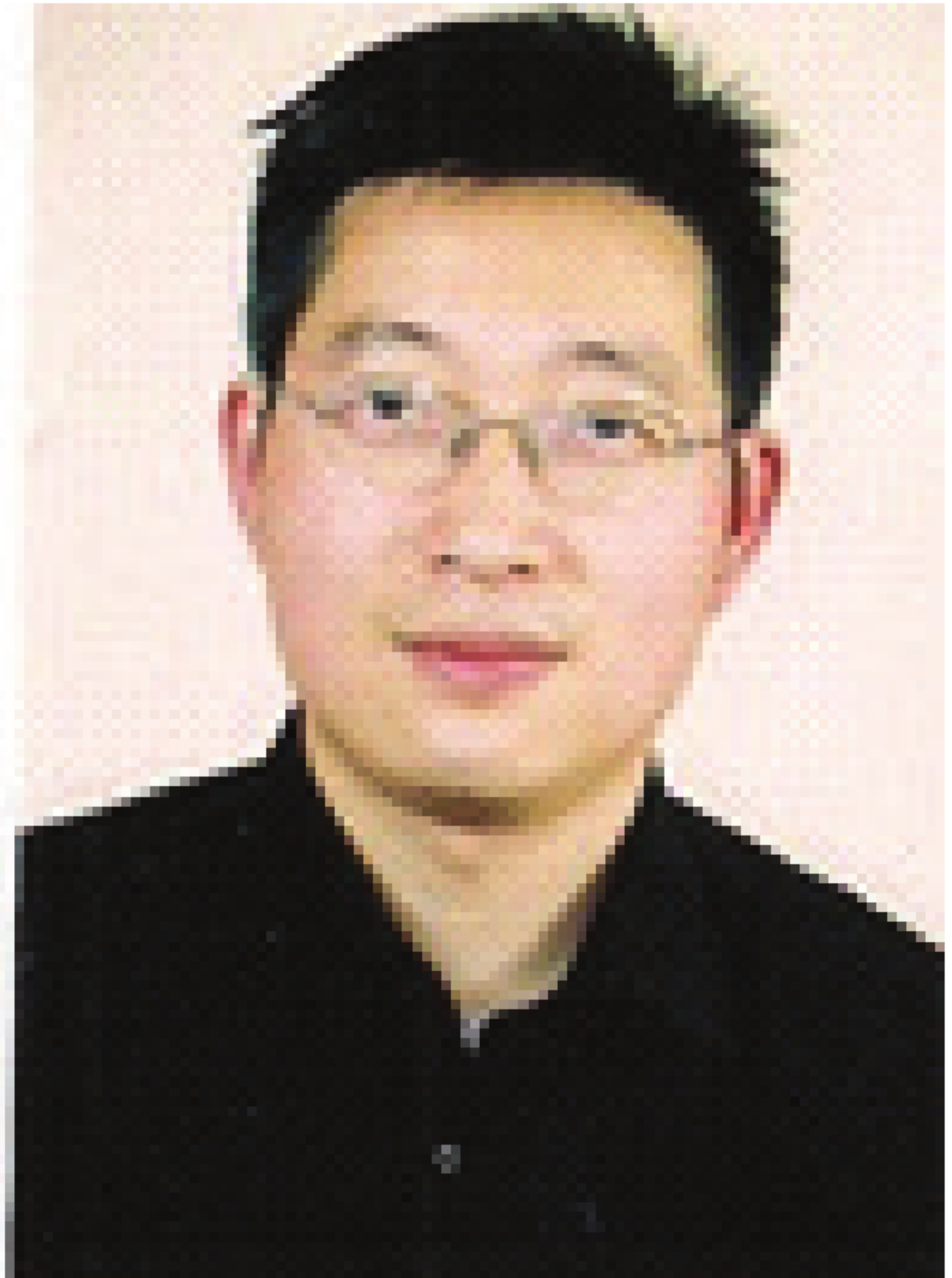}}]{Qihui Wu}
	received the B.S. degree in communications engineering and the M.S. and Ph.D. degrees in communications and information system from the PLA University of Science and Technology, Nanjing, China, in 1994, 1997, and 2000, respectively. He is currently a Professor with the College of Electronic and Information Engineering, Nanjing University of Aeronautics and Astronautics. His current research interests include algorithms and optimization for cognitive wireless networks, soft-defined radio, and wireless communication systems.
\end{IEEEbiography}

\vspace{11pt}

\begin{IEEEbiography}[{\includegraphics[width=1in,height=2.15in,clip,keepaspectratio]{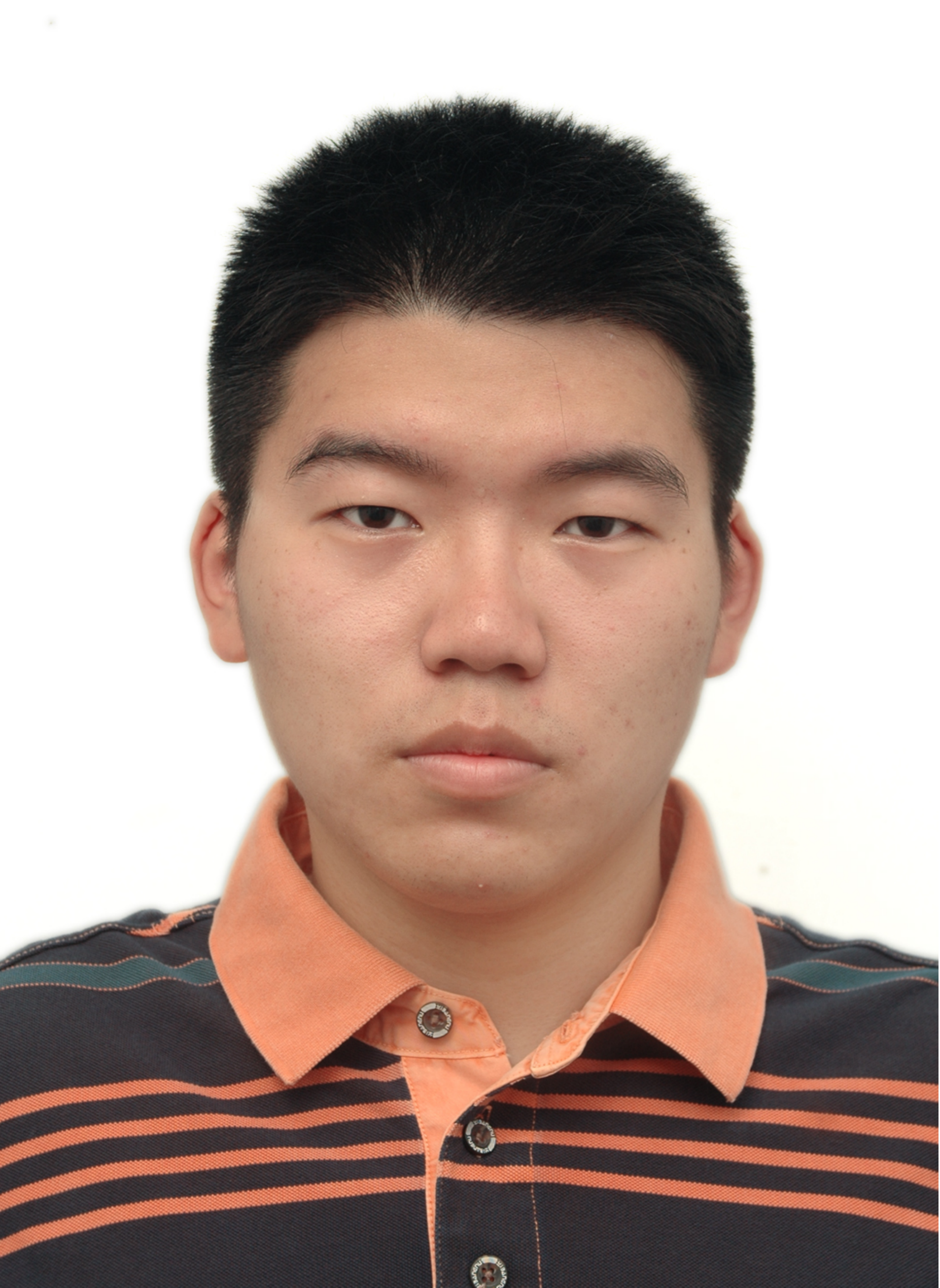}}]{Yang Huang}
	received the B.S. and M. S. degrees from Northeastern University, China, in 2011 and 2013, respectively, and the Ph.D. degree from Imperial College London in 2017. He is currently an Associate Professor with College of Electronic and Information Engineering, Nanjing University of Aeronautics and Astronautics, Nanjing, China. His research interests include wireless communications, MIMO systems, convex optimization, machine learning and signal processing for communications. He has served as Technical Program Committee (TPC) members for many International conferences, such as IEEE GLOBECOM, etc.
\end{IEEEbiography}

\vspace{11pt}

\begin{IEEEbiography}[{\includegraphics[width=1in,height=1.25in,clip,keepaspectratio]{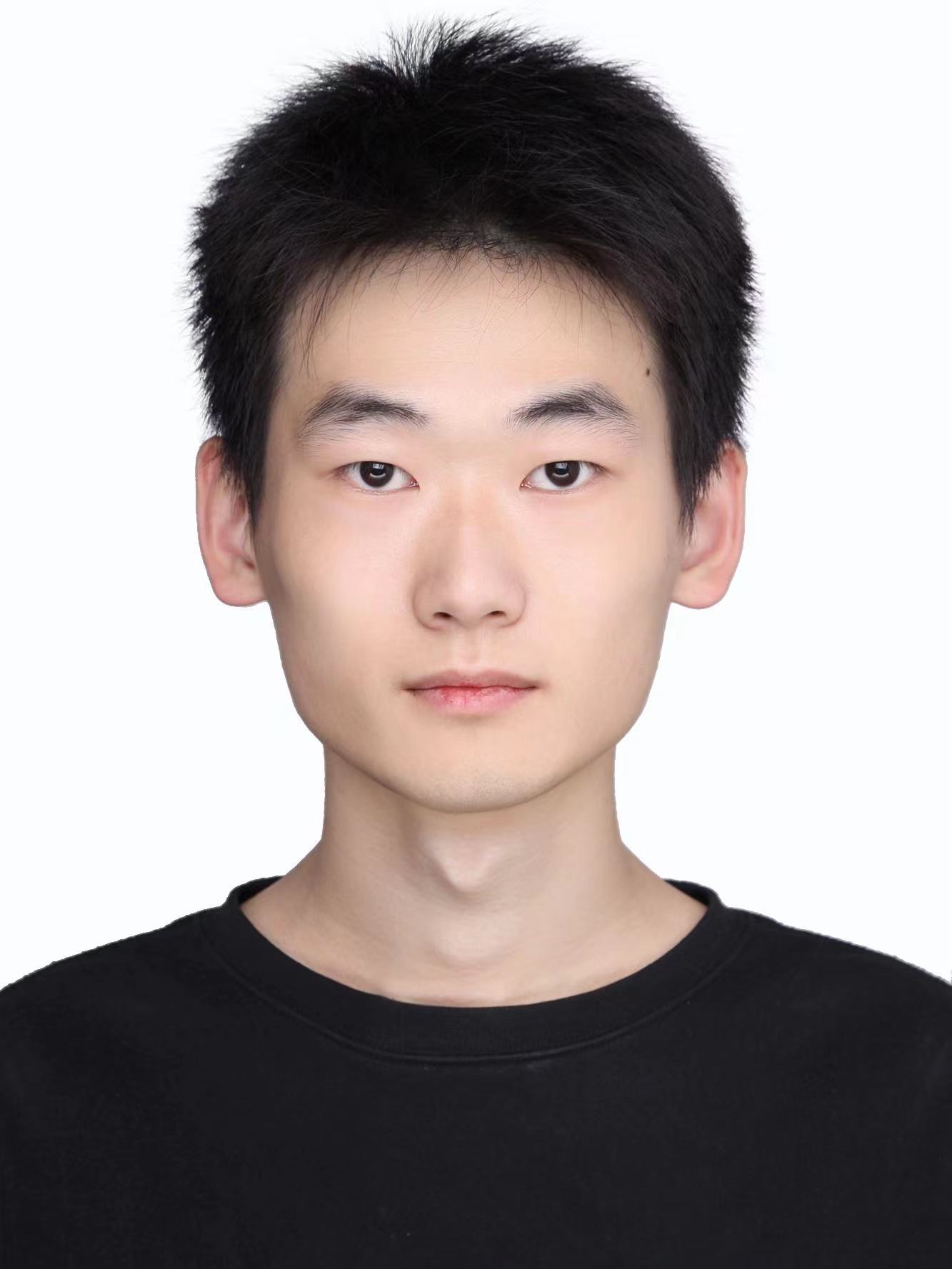}}]{Xuezhao Cai}
	received the B.S. degree in information engineering from the School of Artificial Intelligence, Nanjing University of Information Science and Technology, Nanjing, China , in 2022. He is currently pursuing the M.S. degree in electronic information with the College of Electronic and Information Engineering, Nanjing University of Aeronautics and Astronautics. His current research direction is related to spectrum..
\end{IEEEbiography}

\vspace{11pt}

\begin{IEEEbiography}[{\includegraphics[width=1in,height=1.25in,clip,keepaspectratio]{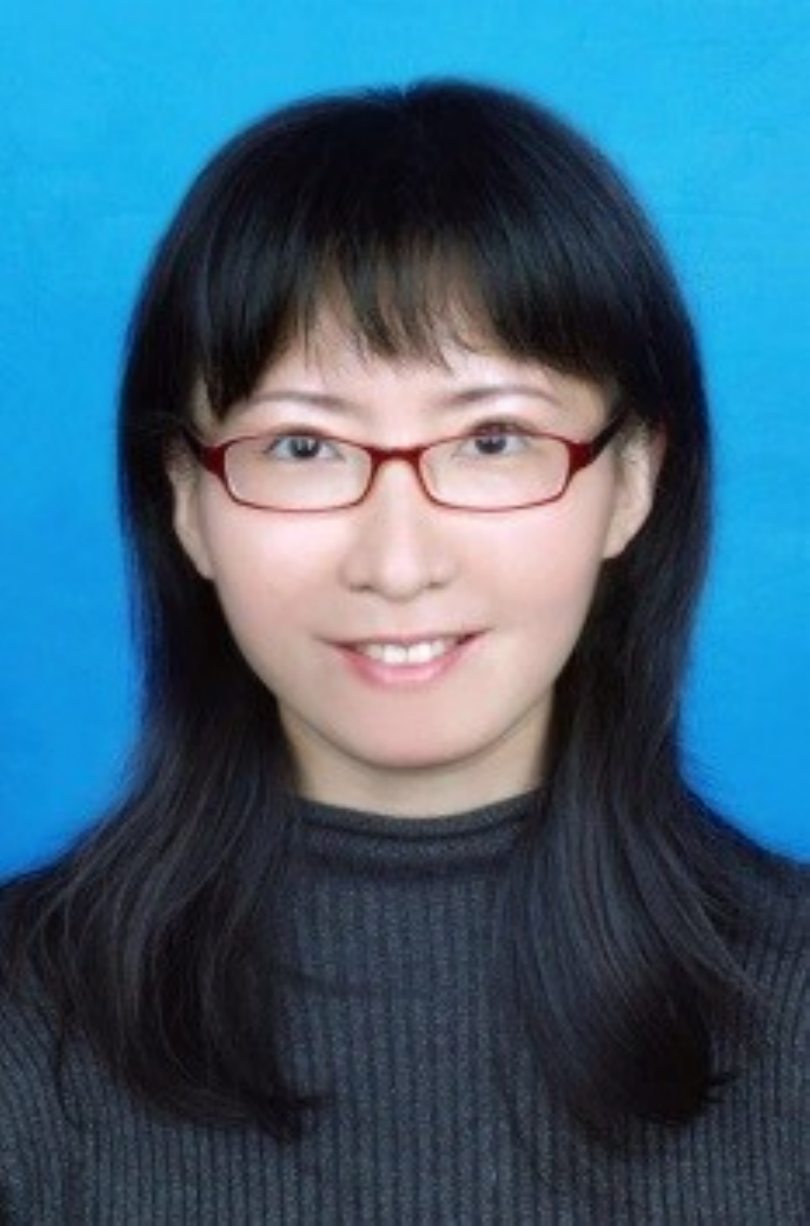}}]{Weizhi Zhong}
	received the B.S. degree in communications engineering and the M.S. and Ph.D. degrees in communications and information system from the PLA University of Science and Technology, Nanjing, China, in 1994, 1997, and 2000, respectively. He is currently a Professor with the College of Electronic and Information Engineering, Nanjing University of Aeronautics and Astronautics. His current research interests include algorithms and optimization for cognitive wireless networks, soft-defined radio, and wireless communication systems.
\end{IEEEbiography}

\vspace{11pt}

\begin{IEEEbiography}[{\includegraphics[width=1in,height=1.25in,clip,keepaspectratio]{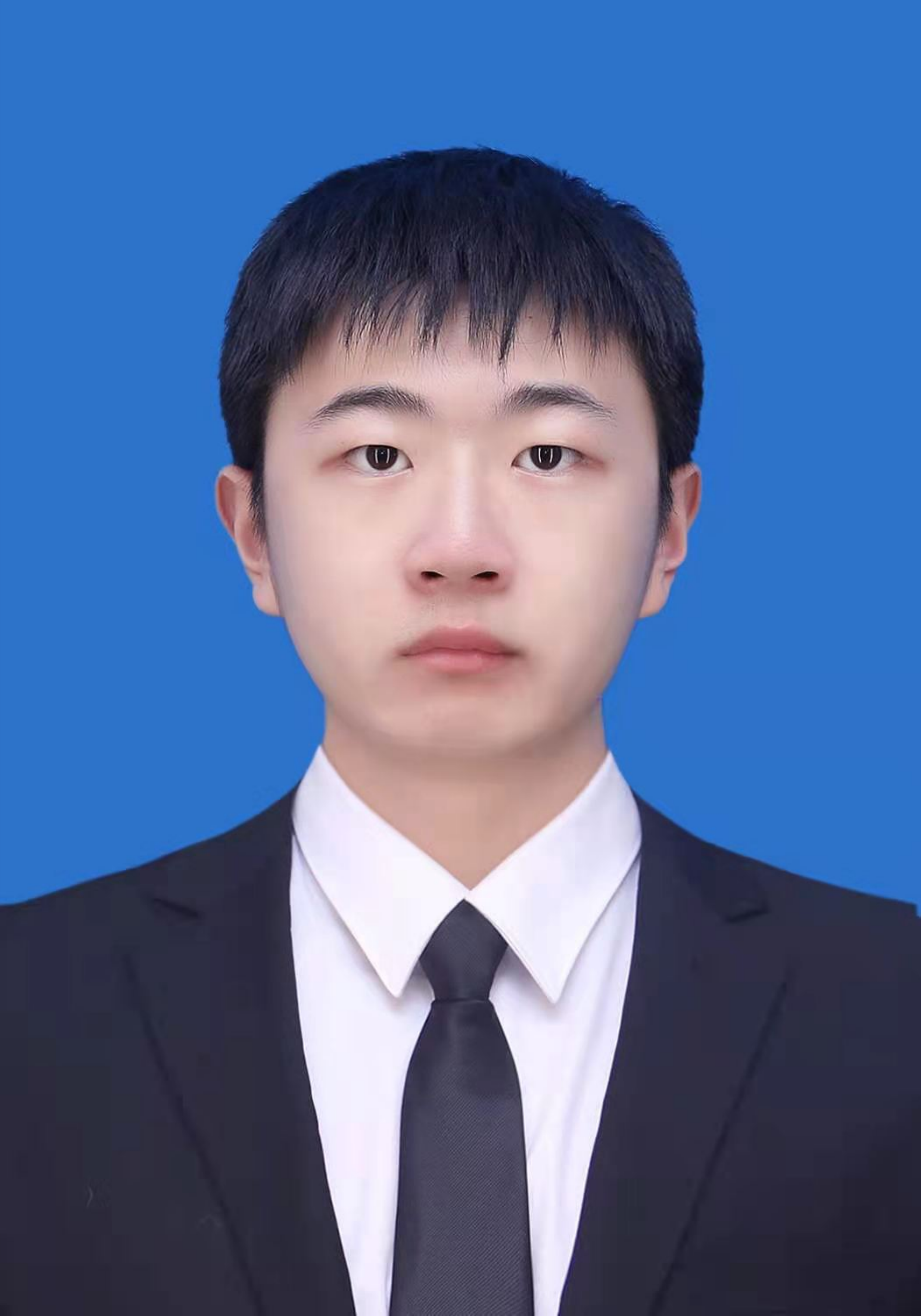}}]{Yi Zhao}
	received the B.S. degree in information engineering from Nanjing University of Aeronautics and Astronautics (NUAA) in 2021. He is currently working towards the master degree in electronic information engineering, NUAA. His current research is temporal and spatial prediction and reconstruction of spectrum situation.
\end{IEEEbiography}

\vfill

\end{document}